
\catcode`@=11 

\font\fourteenrm=cmr10 scaled\magstep2
\font\twelverm=cmr12

\font\ninerm=cmr9

\font\sixrm=cmr6

\font\fourteenbf=cmbx10 scaled\magstep2
\font\twelvebf=cmbx10 scaled\magstep1
\font\ninebf=cmbx9	      \font\sixbf=cmbx6
\font\seventeeni=cmmi10 scaled\magstep3	    \skewchar\seventeeni='177
\font\fourteeni=cmmi10 scaled\magstep2	    \skewchar\fourteeni='177
\font\twelvei=cmmi12                        \skewchar\twelvei='177
\font\ninei=cmmi9			    \skewchar\ninei='177
\font\sixi=cmmi6			    \skewchar\sixi='177
\font\seventeensy=cmsy10 scaled\magstep3    \skewchar\seventeensy='60
\font\fourteensy=cmsy10 scaled\magstep2	    \skewchar\fourteensy='60
\font\twelvesy=cmsy10 scaled\magstep1	    \skewchar\twelvesy='60
\font\ninesy=cmsy9			    \skewchar\ninesy='60
\font\sixsy=cmsy6			    \skewchar\sixsy='60

\font\fourteenex=cmex10 scaled\magstep2
\font\twelveex=cmex10 scaled\magstep1

\font\fourteensl=cmsl10 scaled\magstep2
\font\twelvesl=cmsl10 scaled\magstep1

\font\ninesl=cmsl9

\font\fourteenit=cmti10 scaled\magstep2
\font\twelveit=cmti10 scaled\magstep1
\font\twelvett=cmtt10 scaled\magstep1

%

\font\fourteencp=cmcsc10 scaled\magstep2
\font\twelvecp=cmcsc10 scaled\magstep1
\font\tencp=cmcsc10
\newfam\cpfam
%
%
\newcount\f@ntkey	     \f@ntkey=0
\def\samef@nt{\relax \ifcase\f@ntkey \rm \or\oldstyle \or\or
	 \or\it \or\sl \or\bf \or\tt \or\caps \fi }
%
\jfont\tenmin=min10
\jfont\twelvemin=min10 scaled\magstep1
\jfont\fourteenmin=min10 scaled\magstep2
\jfont\seventeenmin=min10 scaled\magstep3
\jfont\tengt=goth10
\jfont\twelvegt=goth10 scaled\magstep1
\jfont\fourteengt=goth10 scaled\magstep2
\jfont\seventeengt=goth10 scaled\magstep3
\def\fourteenpoint{\relax
    \textfont0=\fourteenrm
    \scriptfont0=\tenrm
    \scriptscriptfont0=\sevenrm
     \def\rm{\fam0 \fourteenrm \f@ntkey=0 }\relax
    \textfont1=\fourteeni	    \scriptfont1=\teni
    \scriptscriptfont1=\seveni
     \def\oldstyle{\fam1 \fourteeni\f@ntkey=1 }\relax
    \textfont2=\fourteensy	    \scriptfont2=\tensy
    \scriptscriptfont2=\sevensy
    \textfont3=\fourteenex     \scriptfont3=\fourteenex
    \scriptscriptfont3=\fourteenex
    \def\it{\fam\itfam \fourteenit\f@ntkey=4 }\textfont\itfam=\fourteenit
    \def\sl{\fam\slfam \fourteensl\f@ntkey=5 }\textfont\slfam=\fourteensl
    \scriptfont\slfam=\tensl
    \def\bf{\fam\bffam \fourteenbf\f@ntkey=6 }\textfont\bffam=\fourteenbf
    \scriptfont\bffam=\tenbf	 \scriptscriptfont\bffam=\sevenbf
    \def\tt{\fam\ttfam \twelvett \f@ntkey=7 }\textfont\ttfam=\twelvett
    \h@big=11.9\p@{} \h@Big=16.1\p@{} \h@bigg=20.3\p@{} \h@Bigg=24.5\p@{}
    \def\caps{\fam\cpfam \twelvecp \f@ntkey=8 }\textfont\cpfam=\twelvecp
    \setbox\strutbox=\hbox{\vrule height 12pt depth 5pt width\z@}
    \samef@nt}
\def\twelvepoint{\relax
    \textfont0=\twelverm
    \scriptfont0=\ninerm
    \scriptscriptfont0=\sixrm
     \def\rm{\fam0 \twelverm \f@ntkey=0 }\relax
    \textfont1=\twelvei            \scriptfont1=\ninei
    \scriptscriptfont1=\sixi
     \def\oldstyle{\fam1 \twelvei\f@ntkey=1 }\relax
    \textfont2=\twelvesy	  \scriptfont2=\ninesy
    \scriptscriptfont2=\sixsy
    \textfont3=\twelveex	  \scriptfont3=\twelveex
    \scriptscriptfont3=\twelveex
    \def\it{\fam\itfam \twelveit \f@ntkey=4 }\textfont\itfam=\twelveit
    \def\sl{\fam\slfam \twelvesl \f@ntkey=5 }\textfont\slfam=\twelvesl
    \scriptfont\slfam=\ninesl
    \def\bf{\fam\bffam \twelvebf \f@ntkey=6 }\textfont\bffam=\twelvebf
    \scriptfont\bffam=\ninebf	  \scriptscriptfont\bffam=\sixbf
    \def\tt{\fam\ttfam \twelvett \f@ntkey=7 }\textfont\ttfam=\twelvett
    \h@big=10.2\p@{}
    \h@Big=13.8\p@{}
    \h@bigg=17.4\p@{}
    \h@Bigg=21.0\p@{}
    \def\caps{\fam\cpfam \twelvecp \f@ntkey=8 }\textfont\cpfam=\twelvecp
    \setbox\strutbox=\hbox{\vrule height 10pt depth 4pt width\z@}
    \samef@nt}
\def\tenpoint{\relax
    \textfont0=\tenrm
    \scriptfont0=\sevenrm
    \scriptscriptfont0=\fiverm
    \def\rm{\fam0 \tenrm \f@ntkey=0 }\relax
    \textfont1=\teni	       \scriptfont1=\seveni
    \scriptscriptfont1=\fivei
    \def\oldstyle{\fam1 \teni \f@ntkey=1 }\relax
    \textfont2=\tensy	       \scriptfont2=\sevensy
    \scriptscriptfont2=\fivesy
    \textfont3=\tenex	       \scriptfont3=\tenex
    \scriptscriptfont3=\tenex
    \def\it{\fam\itfam \tenit \f@ntkey=4 }\textfont\itfam=\tenit
    \def\sl{\fam\slfam \tensl \f@ntkey=5 }\textfont\slfam=\tensl
    \def\bf{\fam\bffam \tenbf \f@ntkey=6 }\textfont\bffam=\tenbf
    \scriptfont\bffam=\sevenbf	   \scriptscriptfont\bffam=\fivebf
    \def\tt{\fam\ttfam \tentt \f@ntkey=7 }\textfont\ttfam=\tentt
    \def\caps{\fam\cpfam \tencp \f@ntkey=8 }\textfont\cpfam=\tencp
    \setbox\strutbox=\hbox{\vrule height 8.5pt depth 3.5pt width\z@}
    \samef@nt}
%
%
%
%
\newdimen\h@big  \h@big=8.5\p@
\newdimen\h@Big  \h@Big=11.5\p@
\newdimen\h@bigg  \h@bigg=14.5\p@
\newdimen\h@Bigg  \h@Bigg=17.5\p@
\def\big#1{{\hbox{$\left#1\vbox to\h@big{}\right.\n@space$}}}
\def\Big#1{{\hbox{$\left#1\vbox to\h@Big{}\right.\n@space$}}}
\def\bigg#1{{\hbox{$\left#1\vbox to\h@bigg{}\right.\n@space$}}}
\def\Bigg#1{{\hbox{$\left#1\vbox to\h@Bigg{}\right.\n@space$}}}
%
%
%
Q4
eLWIcLlpO
X(J
eLWIcLlpO
X(J
\normalbaselineskip = 18pt plus 0.2pt minus 0.1pt            %
\normallineskip = 1.5pt plus 0.1pt minus 0.1pt
\normallineskiplimit = 1.5pt
\newskip\normaldisplayskip
\normaldisplayskip = 20pt plus 5pt minus 10pt
\newskip\normaldispshortskip
\normaldispshortskip = 6pt plus 5pt
\newskip\normalparskip
\normalparskip = 6pt plus 2pt minus 1pt
\newskip\skipregister
\skipregister = 5pt plus 2pt minus 1.5pt
\newif\ifsingl@	   \newif\ifdoubl@
\newif\iftwelv@	   \twelv@true
\def\singlespace{\singl@true\doubl@false\spaces@t}
\def\doublespace{\singl@false\doubl@true\spaces@t}
\def\normalspace{\singl@false\doubl@false\spaces@t}
\def\Tenpoint{\tenpoint\twelv@false\spaces@t}
\def\Twelvepoint{\twelvepoint\twelv@true\spaces@t}
\def\spaces@t{\relax%
 \iftwelv@ \ifsingl@\subspaces@t3:4;\else\subspaces@t1:1;\fi%
 \else \ifsingl@\subspaces@t3:5;\else\subspaces@t4:5;\fi \fi%
 \ifdoubl@ \multiply\baselineskip by 5%
 \divide\baselineskip by 4 \fi \unskip}
\def\subspaces@t#1:#2;{%
      \baselineskip = \normalbaselineskip%
      \multiply\baselineskip by #1 \divide\baselineskip by #2%
      \lineskip = \normallineskip%
      \multiply\lineskip by #1 \divide\lineskip by #2%
      \lineskiplimit = \normallineskiplimit%
      \multiply\lineskiplimit by #1 \divide\lineskiplimit by #2%
      \parskip = \normalparskip%
      \multiply\parskip by #1 \divide\parskip by #2%
      \abovedisplayskip = \normaldisplayskip%
      \multiply\abovedisplayskip by #1 \divide\abovedisplayskip by #2%
      \belowdisplayskip = \abovedisplayskip%
      \abovedisplayshortskip = \normaldispshortskip%
      \multiply\abovedisplayshortskip by #1%
	\divide\abovedisplayshortskip by #2%
      \belowdisplayshortskip = \abovedisplayshortskip%
      \advance\belowdisplayshortskip by \belowdisplayskip%
      \divide\belowdisplayshortskip by 2%
      \smallskipamount = \skipregister%
      \multiply\smallskipamount by #1 \divide\smallskipamount by #2%
      \medskipamount = \smallskipamount \multiply\medskipamount by 2%
      \bigskipamount = \smallskipamount \multiply\bigskipamount by 4 }
\def\normalbaselines{ \baselineskip=\normalbaselineskip%
   \lineskip=\normallineskip \lineskiplimit=\normallineskip%
   \iftwelv@\else \multiply\baselineskip by 4 \divide\baselineskip by 5%
     \multiply\lineskiplimit by 4 \divide\lineskiplimit by 5%
     \multiply\lineskip by 4 \divide\lineskip by 5 \fi }
\Twelvepoint  
\interlinepenalty=50
\interfootnotelinepenalty=5000
\predisplaypenalty=9000
\postdisplaypenalty=500
\hfuzz=1pt
\vfuzz=0.2pt
%
%
%
\def\pagecontents{%
   \ifvoid\topins\else\unvbox\topins\vskip\skip\topins\fi
   \dimen@ = \dp255 \unvbox255
   \ifvoid\footins\else\vskip\skip\footins\footrule\unvbox\footins\fi
   \ifr@ggedbottom \kern-\dimen@ \vfil \fi }
\def\makeheadline{\vbox to 0pt{ \skip@=\topskip
      \advance\skip@ by -12pt \advance\skip@ by -2\normalbaselineskip
      \vskip\skip@ \line{\vbox to 12pt{}\the\headline} \vss
      }\nointerlineskip}
\def\makefootline{\baselineskip = 1.5\normalbaselineskip
		 \line{\the\footline}}
\newif\iffrontpage
\newif\ifletterstyle
\newif\ifp@genum
\def\nopagenumbers{\p@genumfalse}
\def\pagenumbers{\p@genumtrue}
\pagenumbers
\newtoks\paperheadline
\newtoks\letterheadline
\newtoks\letterfrontheadline
\newtoks\lettermainheadline
\newtoks\paperfootline
\newtoks\letterfootline
\newtoks\date
\footline={\ifletterstyle\the\letterfootline\else\the\paperfootline\fi}
\paperfootline={\hss\iffrontpage\else\ifp@genum\tenrm
    -- \folio\ --\hss\fi\fi}
\letterfootline={\hfil}
\headline={\ifletterstyle\the\letterheadline\else\the\paperheadline\fi}
\paperheadline={\hfil}
\letterheadline{\iffrontpage\the\letterfrontheadline
     \else\the\lettermainheadline\fi}
\lettermainheadline={\rm\ifp@genum page \ \folio\fi\hfil\the\date}
\def\monthname{\relax\ifcase\month 0/\or January\or February\or
   March\or April\or May\or June\or July\or August\or September\or
   October\or November\or December\else\number\month/\fi}
\date={\monthname\ \number\day, \number\year}
\countdef\pagenumber=1  \pagenumber=1
\def\advancepageno{\global\advance\pageno by 1
   \ifnum\pagenumber<0 \global\advance\pagenumber by -1
    \else\global\advance\pagenumber by 1 \fi \global\frontpagefalse }
\def\folio{\ifnum\pagenumber<0 \romannumeral-\pagenumber
	   \else \number\pagenumber \fi }
\def\footrule{\dimen@=\prevdepth\nointerlineskip
   \vbox to 0pt{\vskip -0.25\baselineskip \hrule width 0.35\hsize \vss}
   \prevdepth=\dimen@ }
\newtoks\foottokens
\foottokens={\Tenpoint\singlespace}
\newdimen\footindent
\footindent=24pt
\def\vfootnote#1{\insert\footins\bgroup  \the\foottokens
   \interlinepenalty=\interfootnotelinepenalty \floatingpenalty=20000
   \splittopskip=\ht\strutbox \boxmaxdepth=\dp\strutbox
   \leftskip=\footindent \rightskip=\z@skip
   \parindent=0.5\footindent \parfillskip=0pt plus 1fil
   \spaceskip=\z@skip \xspaceskip=\z@skip
   \Textindent{$ #1 $}\footstrut\futurelet\next\fo@t}
\def\Textindent#1{\noindent\llap{#1\enspace}\ignorespaces}
\def\footnote#1{\attach{#1}\vfootnote{#1}}

\let\footsymbol=\star
\newcount\lastf@@t	     \lastf@@t=-1
\newcount\footsymbolcount    \footsymbolcount=0
\newif\ifPhysRev
\def\footsymbolgen{\relax \ifPhysRev \iffrontpage \NPsymbolgen\else
      \PRsymbolgen\fi \else \NPsymbolgen\fi
   \global\lastf@@t=\pageno \footsymbol }
\def\NPsymbolgen{\ifnum\footsymbolcount<0 \global\footsymbolcount=0\fi
   {\iffrontpage \else \advance\lastf@@t by 1 \fi
    \ifnum\lastf@@t<\pageno \global\footsymbolcount=0
     \else \global\advance\footsymbolcount by 1 \fi }
   \ifcase\footsymbolcount \fd@f\star\or \fd@f\dagger\or \fd@f\ast\or
    \fd@f\ddagger\or \fd@f\natural\or \fd@f\diamond\or \fd@f\bullet\or
    \fd@f\nabla\else \fd@f\dagger\global\footsymbolcount=0 \fi }
\def\fd@f#1{\xdef\footsymbol{#1}}
\def\PRsymbolgen{\ifnum\footsymbolcount>0 \global\footsymbolcount=0\fi
      \global\advance\footsymbolcount by -1
      \xdef\footsymbol{\sharp\number-\footsymbolcount} }
\def\space@ver#1{\let\@sf=\empty \ifmmode #1\else \ifhmode
   \edef\@sf{\spacefactor=\the\spacefactor}\unskip${}#1$\relax\fi\fi}
\def\attach#1{\space@ver{\strut^{\mkern 2mu #1} }\@sf\ }
%
%
%
\newcount\chapternumber	     \chapternumber=0
\newcount\sectionnumber	     \sectionnumber=0
\newcount\equanumber	     \equanumber=0
\let\chapterlabel=0
\newtoks\chapterstyle	     \chapterstyle={\Number}
\newskip\chapterskip	     \chapterskip=\bigskipamount
\newskip\sectionskip	     \sectionskip=\medskipamount
\newskip\headskip	     \headskip=8pt plus 3pt minus 3pt
\newdimen\chapterminspace    \chapterminspace=15pc
\newdimen\sectionminspace    \sectionminspace=10pc
\newdimen\referenceminspace  \referenceminspace=25pc
\def\chapterreset{\global\advance\chapternumber by 1
   \ifnum\equanumber<0 \else\global\equanumber=0\fi
   \sectionnumber=0 \makel@bel}
\def\makel@bel{\xdef\chapterlabel{%
\the\chapterstyle{\the\chapternumber}.}}
\def\sectionlabel{\number\sectionnumber \quad }
\def\alphabetic#1{\count255='140 \advance\count255 by #1\char\count255}
\def\Alphabetic#1{\count255='100 \advance\count255 by #1\char\count255}
\def\Roman#1{\uppercase\expandafter{\romannumeral #1}}
\def\roman#1{\romannumeral #1}
\def\Number#1{\number #1}
\def\unnumberedchapters{\let\makel@bel=\relax \let\chapterlabel=\relax
\let\sectionlabel=\relax \equanumber=-1 }
%
\def\titlestyle#1{\par\begingroup \interlinepenalty=9999
     \leftskip=0.03\hsize plus 0.20\hsize minus 0.03\hsize
     \rightskip=\leftskip \parfillskip=0pt
     \hyphenpenalty=9000 \exhyphenpenalty=9000
     \tolerance=9999 \pretolerance=9000
     \spaceskip=0.333em \xspaceskip=0.5em
     \iftwelv@\fourteenpoint\fourteenbf
     \else\twelvepoint\twelvebf\fi
     \noindent  #1\par\endgroup }
\def\spacecheck#1{\dimen@=\pagegoal\advance\dimen@ by -\pagetotal
   \ifdim\dimen@<#1 \ifdim\dimen@>0pt \vfil\break \fi\fi}
\def\chapter#1{\par \penalty-300 \vskip\chapterskip
   \spacecheck\chapterminspace
   \chapterreset \titlestyle{\chapterlabel \ \fourteengt #1}
   \nobreak\vskip\headskip \penalty 30000
   \wlog{\string\chapter\ \chapterlabel } }

%
\def\section#1{\par \ifnum\the\lastpenalty=30000\else
   \penalty-200\vskip\sectionskip \spacecheck\sectionminspace\fi
   \wlog{\string\section\ \chapterlabel\the\sectionnumber}
   \global\advance\sectionnumber by 1  \noindent
   {\caps\enspace\chapterlabel\sectionlabel \twelvegt #1}\par
   \nobreak\vskip\headskip \penalty 30000 }
\def\ssection#1{\par \ifnum\the\lastpenalty=30000\else
   \penalty-200\vskip\sectionskip \spacecheck\sectionminspace\fi
   \wlog{\string\section\ \chapterlabel\the\sectionnumber}
   \global\advance\sectionnumber by 1  \noindent
   {\S \caps\thinspace\chapterlabel\sectionlabel #1}\par
   \nobreak\vskip\headskip \penalty 30000 }
\def\subsection#1{\par
   \ifnum\the\lastpenalty=30000\else \penalty-100\smallskip \fi
   \noindent\undertext{\twelvegt #1}\enspace \vadjust{\penalty5000}}

\def\undertext#1{\vtop{\hbox{#1}\kern 1pt \hrule}}
\def\APPENDIX#1#2{\par\penalty-300\vskip\chapterskip
   \spacecheck\chapterminspace \chapterreset \xdef\chapterlabel {#1}
   \titlestyle{APPENDIX #2} \nobreak\vskip\headskip \penalty 30000
   \wlog{\string\Appendix\ \chapterlabel } }
\def\Appendix#1{\APPENDIX{#1}{#1}}
\def\appendix{\APPENDIX{A}{}}
%
%
%
\def\eqname#1{\relax \ifnum\equanumber<0
     \xdef#1{{\rm(\number-\equanumber)}}\global\advance\equanumber by -1
    \else \global\advance\equanumber by 1
      \xdef#1{{\rm(\chapterlabel \number\equanumber)}} \fi}
\def\eqinsert#1{\noalign{\dimen@=\prevdepth \nointerlineskip
   \setbox0=\hbox to\displaywidth{\hfil #1}
   \vbox to 0pt{\vss\hbox{$\!\box0\!$}\kern-0.5\baselineskip}
   \prevdepth=\dimen@}}
%

%

%

%
%
\def\GENITEM#1;#2{\par \hangafter=0 \hangindent=#1
    \Textindent{#2}\ignorespaces}
\outer\def\newitem#1=#2;{\gdef#1{\GENITEM #2;}}
\newdimen\itemsize		  \itemsize=30pt
\newitem\item=1\itemsize;
\newitem\sitem=1.75\itemsize;	  
\newitem\ssitem=2.5\itemsize;	  
\outer\def\newlist#1=#2&#3&#4;{\toks0={#2}\toks1={#3}%
   \count255=\escapechar \escapechar=-1
   \alloc@0\list\countdef\insc@unt\listcount	 \listcount=0
   \edef#1{\par
      \countdef\listcount=\the\allocationnumber
      \advance\listcount by 1
      \hangafter=0 \hangindent=#4
      \Textindent{\the\toks0{\listcount}\the\toks1}}
   \expandafter\expandafter\expandafter
    \edef\c@t#1{begin}{\par
      \countdef\listcount=\the\allocationnumber \listcount=1
      \hangafter=0 \hangindent=#4
      \Textindent{\the\toks0{\listcount}\the\toks1}}
   \expandafter\expandafter\expandafter
    \edef\c@t#1{con}{\par \hangafter=0 \hangindent=#4 \noindent}
   \escapechar=\count255}
\def\c@t#1#2{\csname\string#1#2\endcsname}
\newlist\point=\Number&.&1.0\itemsize;
\newlist\subpoint=(\alphabetic&)&1.75\itemsize;
\newlist\subsubpoint=(\roman&)&2.5\itemsize;
%

%
%
%
\newcount\referencecount     \referencecount=0
\newif\ifreferenceopen	     \newwrite\referencewrite
\newtoks\rw@toks
\def\NPrefmark#1{\attach{\scriptscriptstyle [ #1 ] }}
\let\PRrefmark=\attach
\def\refmark#1{\relax\ifPhysRev\PRrefmark{#1}\else\NPrefmark{#1}\fi}
\def\refend{\refmark{\number\referencecount}}
\newcount\lastrefsbegincount \lastrefsbegincount=0
\def\refsend{\refmark{\count255=\referencecount
   \advance\count255 by-\lastrefsbegincount
   \ifcase\count255 \number\referencecount
   \or \number\lastrefsbegincount,\number\referencecount
   \else \number\lastrefsbegincount-\number\referencecount \fi}}
\def\refch@ck{\chardef\rw@write=\referencewrite
   \ifreferenceopen \else \referenceopentrue
   \immediate\openout\referencewrite=reference.aux \fi}
%
{\catcode`\^^M=\active 
  \gdef\obeyendofline{\catcode`\^^M\active \let^^M\ }}%
%
{\catcode`\^^M=\active 
  \gdef\ignoreendofline{\catcode`\^^M=5}}
{\obeyendofline\gdef\rw@start#1{\def\t@st{#1} \ifx\t@st\blankend%
\endgroup \@sf \relax \else \ifx\t@st\bl@nkend \endgroup \@sf \relax%
\else \rw@begin#1
\backtotext
\fi \fi } }
{\obeyendofline\gdef\rw@begin#1
{\def\n@xt{#1}\rw@toks={#1}\relax%
\rw@next}}
\def\blankend{}
{\obeylines\gdef\bl@nkend{
}}
\newif\iffirstrefline  \firstreflinetrue
\def\rwr@teswitch{\ifx\n@xt\blankend \let\n@xt=\rw@begin %
 \else\iffirstrefline \global\firstreflinefalse%
\immediate\write\rw@write{\noexpand\obeyendofline \the\rw@toks}%
\let\n@xt=\rw@begin%
      \else\ifx\n@xt\rw@@d \def\n@xt{\immediate\write\rw@write{%
	\noexpand\ignoreendofline}\endgroup \@sf}%
	     \else \immediate\write\rw@write{\the\rw@toks}%
	     \let\n@xt=\rw@begin\fi\fi \fi}
\def\rw@next{\rwr@teswitch\n@xt}
\def\rw@@d{\backtotext} \let\rw@end=\relax
\let\backtotext=\relax

\newdimen\refindent	\refindent=30pt
\def\refitem#1{\par \hangafter=0 \hangindent=\refindent \Textindent{#1}}
\def\REFNUM#1{\space@ver{}\refch@ck \firstreflinetrue%
 \global\advance\referencecount by 1 \xdef#1{\the\referencecount}}
\def\refnum#1{\space@ver{}\refch@ck \firstreflinetrue%
 \global\advance\referencecount by 1 \xdef#1{\the\referencecount}\refend}

\def\REF#1{\REFNUM#1%
 \immediate\write\referencewrite{%
 \noexpand\refitem{#1.}}%
\begingroup\obeyendofline\rw@start}
\def\ref{\refnum\?%
 \immediate\write\referencewrite{\noexpand\refitem{\?.}}%
\begingroup\obeyendofline\rw@start}
\def\Ref#1{\refnum#1%
 \immediate\write\referencewrite{\noexpand\refitem{#1.}}%
\begingroup\obeyendofline\rw@start}
\def\REFS#1{\REFNUM#1\global\lastrefsbegincount=\referencecount
\immediate\write\referencewrite{\noexpand\refitem{#1.}}%
\begingroup\obeyendofline\rw@start}
\newcount\figurecount	  \figurecount=0
\newif\iffigureopen	  \newwrite\figurewrite
\def\figch@ck{\chardef\rw@write=\figurewrite \iffigureopen\else
   \immediate\openout\figurewrite=figures.aux
   \figureopentrue\fi}
\def\FIGNUM#1{\space@ver{}\figch@ck \firstreflinetrue%
 \global\advance\figurecount by 1 \xdef#1{\the\figurecount}}
\def\FIG#1{\FIGNUM#1
   \immediate\write\figurewrite{\noexpand\refitem{#1.}}%
   \begingroup\obeyendofline\rw@start}
\def\fig{\FIGNUM\? fig.~\?%
\immediate\write\figurewrite{\noexpand\refitem{\?.}}%
\begingroup\obeyendofline\rw@start}
\def\figure{\FIGNUM\? figure~\?
   \immediate\write\figurewrite{\noexpand\refitem{\?.}}%
   \begingroup\obeyendofline\rw@start}
\def\Fig{\FIGNUM\? Fig.~\?%
\immediate\write\figurewrite{\noexpand\refitem{\?.}}%
\begingroup\obeyendofline\rw@start}
\def\Figure{\FIGNUM\? Figure~\?%
\immediate\write\figurewrite{\noexpand\refitem{\?.}}%
\begingroup\obeyendofline\rw@start}
\newcount\tablecount	 \tablecount=0
\newif\iftableopen	 \newwrite\tablewrite
\def\tabch@ck{\chardef\rw@write=\tablewrite \iftableopen\else
   \immediate\openout\tablewrite=tables.aux
   \tableopentrue\fi}
\def\TABNUM#1{\space@ver{}\tabch@ck \firstreflinetrue%
 \global\advance\tablecount by 1 \xdef#1{\the\tablecount}}
\def\TABLE#1{\TABNUM#1
   \immediate\write\tablewrite{\noexpand\refitem{#1.}}%
   \begingroup\obeyendofline\rw@start}
\def\Table{\TABNUM\? Table~\?%
\immediate\write\tablewrite{\noexpand\refitem{\?.}}%
\begingroup\obeyendofline\rw@start}
%

%
%
%
\def\masterreset{\global\pagenumber=1 \global\chapternumber=0
   \global\equanumber=0 \global\sectionnumber=0
   \global\referencecount=0 \global\figurecount=0 \global\tablecount=0 }
\def\FRONTPAGE{\ifvoid255\else\vfill\penalty-2000\fi
      \masterreset\global\frontpagetrue
      \global\lastf@@t=0 \global\footsymbolcount=0}

\def\paperstyle{\letterstylefalse\normalspace\papersize}
\def\letterstyle{\letterstyletrue\singlespace\lettersize}
%
\def\papersize{\hsize=35.2pc\vsize=52.7pc\hoffset=0.5pc\voffset=0.8pc
	       \skip\footins=\bigskipamount}
\def\lettersize{\hsize=35.2pc\vsize=50.0pc\hoffset=0.5pc\voffset=2.5pc
   \skip\footins=\smallskipamount \multiply\skip\footins by 3 }
\paperstyle   
%
%
\def\MEMO{\letterstyle\FRONTPAGE \letterfrontheadline={\hfil}
    \line{\quad\fourteenrm NU MEMORANDUM\hfil\twelverm\the\date\quad}
    \medskip \memod@f}

\def\memit@m#1{\smallskip \hangafter=0 \hangindent=1in
      \Textindent{\caps #1}}
\def\memod@f{\xdef\to{\memit@m{To:}}\xdef\from{\memit@m{From:}}%
     \xdef\topic{\memit@m{Topic:}}\xdef\subject{\memit@m{Subject:}}%
     \xdef\rule{\bigskip\hrule height 1pt\bigskip}}
\memod@f
%


%

%

%

%
\newskip\lettertopfil
\lettertopfil = -40pt plus 10pt minus 10pt
\newskip\letterbottomfil
\letterbottomfil = 0pt plus 2.3in minus 0pt
\newskip\spskip \setbox0\hbox{\ } \spskip=-1\wd0
\def\addressee#1{\medskip \rightline{\the\date\hskip 30pt} \bigskip
   \vskip\lettertopfil
   \ialign to\hsize{\strut ##\hfil\tabskip 0pt plus \hsize \cr #1 \crcr}
   \medskip\noindent\hskip\spskip}
\newskip\signatureskip	     \signatureskip=50pt
\def\signed#1{\par \penalty 9000 \bigskip \dt@pfalse
  \everycr={\noalign{\ifdt@p\vskip\signatureskip\global\dt@pfalse\fi}}
  \setbox0=\vbox{\singlespace \halign{\tabskip 0pt \strut ##\hfil\cr
   \noalign{\global\dt@ptrue}#1\crcr}}
  \line{\hskip 0.5\hsize minus 0.5\hsize \box0\hfil} \medskip }

\def\endletter{\ifnum\pagenumber=1 \vskip\letterbottomfil\supereject
\else \vfil\supereject \fi}
\newbox\letterb@x
\def\lettertext{\par\unvcopy\letterb@x\par}
\def\multiletter{\setbox\letterb@x=\vbox\bgroup
      \everypar{\vrule height 1\baselineskip depth 0pt width 0pt }
      \singlespace \topskip=\baselineskip }
\def\letterend{\par\egroup}
%
%
%
\newskip\frontpageskip
\newtoks\pubtype
\newtoks\Pubnum
\newtoks\pubnum
\newif\ifp@bblock  \p@bblocktrue
\def\PH@SR@V{\doubl@true \baselineskip=24.1pt plus 0.2pt minus 0.1pt
	     \parskip= 3pt plus 2pt minus 1pt }
\def\PHYSREV{\paperstyle\PhysRevtrue\PH@SR@V}
%
\newif\ifYUKAWA  \YUKAWAfalse
\def\YUKAWAmark{\hbox{\elevenmib
   Yukawa\hskip0.05cm Hall\hskip0.05cm Kyoto \hfill}}
\def\titlepage{\FRONTPAGE\paperstyle\ifPhysRev\PH@SR@V\fi
    \ifYUKAWA\null\vskip-1.70cm\YUKAWAmark\vskip0.6cm\fi
    \ifp@bblock\p@bblock\fi}
\def\nopubblock{\p@bblockfalse}
\def\endpage{\vfil\break}
\frontpageskip=1\medskipamount plus .5fil
%
\pubtype={ }
\newtoks\publevel
\publevel={Report}   
\Pubnum={\the\pubnum}
\pubnum={    }
%
\def\p@bblock{\begingroup \tabskip=\hsize minus \hsize
   \baselineskip=1.5\ht\strutbox \topspace-2\baselineskip
   \halign to\hsize{\strut ##\hfil\tabskip=0pt\crcr
   \the\Pubnum\cr \the\date\cr }\endgroup}
%
\def\title#1{\vskip\frontpageskip\vfill
   {\fourteenbf\seventeenmin\titlestyle{#1}}\vskip\headskip\vfill }
\def\author#1{\vskip\frontpageskip\titlestyle{\twelvemin\twelvecp #1}\nobreak}

%
\def\address#1{\par\kern 5pt \titlestyle{\twelvepoint\sl #1}}
\def\andaddress{\par\kern 5pt \centerline{\sl and} \address}
%
\def\abstract#1{\vfill\vskip\frontpageskip
    \hbox{{\fourteencp Abstract}\hfill}\vskip\headskip#1\vfill\endpage}

%

%
%
%

\def\\{\relax\ifmmode\backslash\else$\backslash$\fi}
\def\globaleqnumbers{\relax\if\equanumber<0\else\global\equanumber=-1\fi}

\def\journal#1&#2(#3){\unskip, \sl #1~\bf #2 \rm (19#3) }

\def\topspace{\hrule height 0pt depth 0pt \vskip}

\let\int=\intop		
\def\prop{\mathrel{{\mathchoice{\pr@p\scriptstyle}{\pr@p\scriptstyle}{
		\pr@p\scriptscriptstyle}{\pr@p\scriptscriptstyle} }}}
\def\pr@p#1{\setbox0=\hbox{$\cal #1 \char'103$}
   \hbox{$\cal #1 \char'117$\kern-.4\wd0\box0}}
\def\lsim{\mathrel{\mathpalette\@versim<}}
\def\gsim{\mathrel{\mathpalette\@versim>}}
\def\@versim#1#2{\lower0.2ex\vbox{\baselineskip\z@skip\lineskip\z@skip
  \lineskiplimit\z@\ialign{$\m@th#1\hfil##\hfil$\crcr#2\crcr\sim\crcr}}}
%
%
%
\let\sec@nt=\sec
\def\sec{\relax\ifmmode\let\n@xt=\sec@nt\else\let\n@xt\section\fi\n@xt}
\def\obsolete#1{\message{Macro \string #1 is obsolete.}}
\def\firstsec#1{\obsolete\firstsec \section{#1}}
\def\firstsubsec#1{\obsolete\firstsubsec \subsection{#1}}
\def\thispage#1{\obsolete\thispage \global\pagenumber=#1\frontpagefalse}
\def\thischapter#1{\obsolete\thischapter \global\chapternumber=#1}
\def\nextequation#1{\obsolete\nextequation \global\equanumber=#1
   \ifnum\the\equanumber>0 \global\advance\equanumber by 1 \fi}
\def\BOXITEM{\afterassigment\B@XITEM\setbox0=}
\def\B@XITEM{\par\hangindent\wd0 \noindent\box0 }
%

%
\catcode`@=12 
\message{ by V.K.}
\relax

\def\yen{\hbox{Y\kern-0.75em =}}
%
%
\catcode`@=11
%
%
\font\fourteenmib=cmmib10 scaled\magstep2    \skewchar\fourteenmib='177
\font\twelvemib=cmmib10 scaled\magstep1	    \skewchar\twelvemib='177
\font\elevenmib=cmmib10 scaled\magstephalf   \skewchar\elevenmib='177
\font\tenmib=cmmib10			    \skewchar\tenmib='177
%
\font\fourteenbsy=cmbsy10 scaled\magstep2     \skewchar\fourteenbsy='60
\font\twelvebsy=cmbsy10 scaled\magstep1	      \skewchar\twelvebsy='60
\font\elevenbsy=cmbsy10 scaled\magstephalf    \skewchar\elevenbsy='60
\font\tenbsy=cmbsy10			      \skewchar\tenbsy='60
%
\newfam\mibfam
\def\samef@nt{\relax \ifcase\f@ntkey \rm \or\oldstyle \or\or
	 \or\it \or\sl \or\bf \or\tt \or\caps \or\mib \fi }
\def\fourteenpoint{\relax
    \textfont0=\fourteenrm	
    \scriptfont0=\tenrm
    \scriptscriptfont0=\sevenrm
     \def\rm{\fam0 \fourteenrm \f@ntkey=0 }\relax
    \textfont1=\fourteeni	    \scriptfont1=\teni
    \scriptscriptfont1=\seveni
     \def\oldstyle{\fam1 \fourteeni\f@ntkey=1 }\relax
    \textfont2=\fourteensy	    \scriptfont2=\tensy
    \scriptscriptfont2=\sevensy
    \textfont3=\fourteenex     \scriptfont3=\fourteenex
    \scriptscriptfont3=\fourteenex
    \def\it{\fam\itfam \fourteenit\f@ntkey=4 }\textfont\itfam=\fourteenit
    \def\sl{\fam\slfam \fourteensl\f@ntkey=5 }\textfont\slfam=\fourteensl
    \scriptfont\slfam=\tensl
    \def\bf{\fam\bffam \fourteenbf\f@ntkey=6 }\textfont\bffam=\fourteenbf
    \scriptfont\bffam=\tenbf	 \scriptscriptfont\bffam=\sevenbf
    \def\tt{\fam\ttfam \twelvett \f@ntkey=7 }\textfont\ttfam=\twelvett
    \h@big=11.9\p@{} \h@Big=16.1\p@{} \h@bigg=20.3\p@{} \h@Bigg=24.5\p@{}
    \def\caps{\fam\cpfam \twelvecp \f@ntkey=8 }\textfont\cpfam=\twelvecp
    \setbox\strutbox=\hbox{\vrule height 12pt depth 5pt width\z@}
    \def\mib{\fam\mibfam \fourteenmib \f@ntkey=9 }
    \textfont\mibfam=\fourteenmib      \scriptfont\mibfam=\tenmib
    \scriptscriptfont\mibfam=\tenmib
    \samef@nt}
\def\twelvepoint{\relax
    \textfont0=\twelverm
    \scriptfont0=\ninerm
    \scriptscriptfont0=\sixrm
     \def\rm{\fam0 \twelverm \f@ntkey=0 }\relax
     \textfont1=\twelvei          \scriptfont1=\ninei
    \scriptscriptfont1=\sixi
     \def\oldstyle{\fam1 \twelvei\f@ntkey=1 }\relax
    \textfont2=\twelvesy	  \scriptfont2=\ninesy
    \scriptscriptfont2=\sixsy
    \textfont3=\twelveex	  \scriptfont3=\twelveex
    \scriptscriptfont3=\twelveex
    \def\it{\fam\itfam \twelveit \f@ntkey=4 }\textfont\itfam=\twelveit
    \def\sl{\fam\slfam \twelvesl \f@ntkey=5 }\textfont\slfam=\twelvesl
    \scriptfont\slfam=\ninesl
    \def\bf{\fam\bffam \twelvebf \f@ntkey=6 }\textfont\bffam=\twelvebf
    \scriptfont\bffam=\ninebf	  \scriptscriptfont\bffam=\sixbf
    \def\tt{\fam\ttfam \twelvett \f@ntkey=7 }\textfont\ttfam=\twelvett
    \h@big=10.2\p@{}
    \h@Big=13.8\p@{}
    \h@bigg=17.4\p@{}
    \h@Bigg=21.0\p@{}
    \def\caps{\fam\cpfam \twelvecp \f@ntkey=8 }\textfont\cpfam=\twelvecp
    \setbox\strutbox=\hbox{\vrule height 10pt depth 4pt width\z@}
    \def\mib{\fam\mibfam \twelvemib \f@ntkey=9 }
    \textfont\mibfam=\twelvemib	    \scriptfont\mibfam=\tenmib
    \scriptscriptfont\mibfam=\tenmib
    \samef@nt}
\def\tenpoint{\relax
    \textfont0=\tenrm
    \scriptfont0=\sevenrm
    \scriptscriptfont0=\fiverm
    \def\rm{\fam0 \tenrm \f@ntkey=0 }\relax
    \textfont1=\teni	       \scriptfont1=\seveni
    \scriptscriptfont1=\fivei
    \def\oldstyle{\fam1 \teni \f@ntkey=1 }\relax
    \textfont2=\tensy	       \scriptfont2=\sevensy
    \scriptscriptfont2=\fivesy
    \textfont3=\tenex	       \scriptfont3=\tenex
    \scriptscriptfont3=\tenex
    \def\it{\fam\itfam \tenit \f@ntkey=4 }\textfont\itfam=\tenit
    \def\sl{\fam\slfam \tensl \f@ntkey=5 }\textfont\slfam=\tensl
    \def\bf{\fam\bffam \tenbf \f@ntkey=6 }\textfont\bffam=\tenbf
    \scriptfont\bffam=\sevenbf	   \scriptscriptfont\bffam=\fivebf
    \def\tt{\fam\ttfam \tentt \f@ntkey=7 }\textfont\ttfam=\tentt
    \def\caps{\fam\cpfam \tencp \f@ntkey=8 }\textfont\cpfam=\tencp
    \setbox\strutbox=\hbox{\vrule height 8.5pt depth 3.5pt width\z@}
    \def\mib{\fam\mibfam \tenmib \f@ntkey=9 }
    \textfont\mibfam=\tenmib   \scriptfont\mibfam=\tenmib
    \scriptscriptfont\mibfam=\tenmib
    \samef@nt}
%
%
\Twelvepoint
\catcode`@=12
%
%
%
\unnumberedchapters
\def\under#1{$\underline{\raise4pt{\hbox{#1}}}$}
\def\delbvec{\partial^{\raise3pt\hbox{$\!\!\!\!\leftrightarrow$}}}
\def\ymh{Yang-Mills-Higgs lagrangian$\;$}

\def\ws{Weinberg-Salam theory$\;$}
\def\dis{\displaystyle}
\def\={\,=\,}
\def\+{\,+\,}
\def\-{\,-\,}
\def\ii{I\hskip -0.3mm I}
\def\iii{I\hskip -0.4mm I\hskip -0.4mm I}
\def\cham{Chamseddine et al.$\;$}

\titlepage
\title
{
 $B@@@(JGauge theory and Higgs mechanism $B@@@(Jbased on differential
geometry
 on discrete space $M_4\times Z_{\mathop{}_{N}}$
}
\author{Yoshitaka Okumura}
\address{\hskip -0.8cm
  Department of Applied Physics,
  Chubu University, Kasugai, Aichi, 487
}
\vskip 2cm
\abstract{
\hskip -2mm
\ws and $SU(5)$ grand unified theory \hskip 1mm (Gut)\hskip 2mm
are reconstructed using
the generalized differential calculus
extended on the discrete space $M_4\times Z_{\mathop{}_{N}}$. Our starting
point is the generalized gauge field expressed by
$A(x,n)=\!\sum_{i}a^\dagger_{i}(x,n){\bf d}a_i(x,n), (n=1,2,\cdots N)$,
where $a_i(x,n)$
is the square matrix valued function defined on $M_4\times Z_{\mathop{}_{N}}$
and ${\bf d}=d+\sum_{m=1}^{\mathop{}_{N}}d_{\chi_m}$ is generalized
exterior derivative. We can construct the consistent algebra of $d_{\chi_m}$
which is exterior derivative with respect to $Z_{\mathop{}_{N}}$ and the
spontaneous breakdown of gauge symmetry is coded in ${d_{\chi_m}}$.
The unified picture of the gauge field and Higgs field as the generalized
connection in non-commutative geometry is realized.
Not only \ymh but also Dirac lagrangian, invariant against the gauge
transformation, are reproduced through the inner product between the
differential forms.
Three sheets ($Z_3)$ are necessary for \ws including strong interaction
and $SU(5)$ Gut.
Our formalism is applicable to more realistic model
like $SO(10)$ unification model.
}

\section{
{
\hskip 5.7cm
{\bf I. Introduction}
}}
Higgs mechanism is inevitable to probe the spontaneously broken gauge theory.
\ws[1] and other physically interesting model [2]
are all based on this mechanism and have elucidated many physical mysteries
in particle physics. However, it seems to be rather artificial and Higgs
particle as the essential ingredient of this mechanism has not been detected
from now, which have persuaded one to investigate other alternatives such as
the technicolor model, top quark condensation [3] and etc.. Such alternatives
, however, are by no means successful and satisfactory because of the extra
physical modes like the technicolor particles. \par
Connes [4] has proposed new approach based on the non-commutative geometry
in the discrete space composed of a continuous manifold $M_4$ and a discrete
two point space $Z_2$ which enables one to regard the Higgs field
as a kind of the gauge field. Connes idea is attractive also in respect of
non-introduction of the extra physical freedoms. However, his theory is
mathematically difficult so that one is hard to understand it. Several versions
have been extended according to Connes's philosophy to make this approach
more understandable [5]. The version of Chamseddine, Felder and Fr{\"o}lich
 [6] is most successful and elaborate among them.
Inclusion of the strong interaction
is more difficult in Connes's original version. \cham provide the formalism
to be able to treat not only \ws but also more complex model like $SU(5)$
grand unified theory (Gut) by enlarging the discrete space to $M_4\times
Z_{\mathop{}_{N}}$.
Their approach
is essentially in line with Connes' version. \par
Sitarz [7] proposed the generalized differential calculus on $M_4\times Z_2$
to realize the unified picture of gauge field and Higgs field. It looks
fairly different from that of \cham. He introduced the fifth one form basis
 $\chi$ and the exterior derivative $d_\chi$, characterizing the discrete
 direction $Z_2$ which are on the same footing as the ordinary one form basis
  $dx^\mu$ and the exterior derivative $d$.
  \par
  The present author and K. Morita
[8] developed this approach adopting the algebraic rules different from
that of Sitarz also on $M_4\times Z_2$. We introduced the matrix function
$M(y)$ ( $y$ is a variable in $Z_2$ and $y=\pm$ ) which is a factor
 to determine the scale and pattern of the spontaneous breakdown of
 gauge symmetry. The introduction of $M(y)$ makes the formalism more flexible.
 However, the gauge non-invariant term appears in Higgs potential and
 we have to discard it by hand. This defect in our first paper [8]
  has been overcome by taking account of the auxiliary fields according to
  \cham [6]  and we could
  reproduce \ws in the completely gauge invariant way [9]. Its generalization
  to $M_4\times Z_{\mathop{}_{N}}$ has been done by the present author and K.
Morita [10]
  and $SU(5)$ Gut is reconstructed there.
  However, the inclusion of the strong interaction into \ws  and
  the incorporation of fermion sector in $SU(5)$ Gut
  were not successful. \par
The purpose of this paper is to present the consistent formalism to construct
the spontaneously broken gauge theories like \ws and $SU(5)$ Gut
based on the non-commutative geometry in the discrete space $M_4\times
Z_{\mathop{}_{N}}$.
This paper is divided into six sections. The next section presents
the general framework based on the generalized differential calculus on
$M_4\times Z_{\mathop{}_{N}}$. The generalized gauge field is defined there and
a geometrical
picture for the unification of the gauge and Higgs fields is realized.
The third section provides the gauge invariant lagrangian for both bosons
 and fermions on the same footing. The fourth section is devoted to the
 reconstruction of \ws. Three sheets ($N=3$) are necessary for the inclusion
 of strong interaction. $SU(5)$ Gut is treated in detail in sec.V where
 the fermionic sector is also incorporated. The last section is devoted for
 conclusions and discussions.

\section{
{
\hskip 2.2cm
{\bf \ii. Generalized gauge field defined on $M_4\times Z_{\mathop{}_{N}}$}
}}
Let $ A(x,n)$ be the generalized gauge field defined on the discrete space
$M_4 \times Z_{\mathop{}_{N}}$, where $M_4$ is the ordinary
four dimensional Minkowski
space and $Z_{\mathop{}_{N}}$ is the space with the discrete variable  $n\,
(n=1,2,\cdots N) $. $A(x,n)$ is
the square-matrix-valued one-form denoted by
$$
      A(x,n)\=\sum_{i}a^\dagger_{i}(x,n){\bf d}a_i(x,n),\eqno(1)
$$
where $a_i(x,n)$ is also the square-matrix-valued function
and ${\bf d}$ is the generalized exterior derivative defined as follows.
$$
       \eqalign{{\bf d}a_{i}(x,n)&\=(d + d_\chi)a(x,n)
                                   =(d+\sum_{k=1}^{\mathop{}_{N}}
                                     d_{\chi_k})a(x,n) \cr
     da_i(x,n) &\= \partial_\mu a_i(x,n)dx^\mu,\hskip 1cm
     \partial_\mu \equiv {\partial \over \partial x^\mu},\cr
   d_{\chi_k} a_i(x,n) &\= [-a_i(x,n)M_{nk}\chi_k + M_{nk}\chi_ka_i(x,n)]. \cr
                }
        \eqno(2)
$$
\vfill\eject
Here $dx^\mu$ is
ordinary one form basis, taken to be dimensionless, in $M_4$, and $\chi_k$
is the one form basis in the discrete space $Z_{\mathop{}_{N}}$ assumed to be
also dimensionless,
 and we assume $\chi_k^\dagger\=-\chi_k$.
We have introduced $x$-independent matrix $M_{nm}$ whose hermitian conjugation
is given by $M_{nk}^\dagger\=M_{kn}$. The matrix $M_{nk}$ turns out to
determine the scale and pattern of the spontaneous breakdown of the gauge
symmetry.\par
We then investigate the condition that the Leibniz rule with respect to ${\bf
d}$ consistently holds. The Leibniz rule can be written as (the subscript $i$
of
$a_i(x,n)$ is abbreviated for a while)
$$
  {\bf d}(a(x,n)b(x,n))=({\bf d}a(x,n))b(x,n)+a(x,n)({\bf d}b(x,n)), \eqno(3a)
$$
where the $b(x,n)$ is also the matrix valued function like $a(x,n)$.
It is evident that
$$
   d(a(x,n)b(x,n))=(d a(x,n))b(x,n)+a(x,n)( d b(x,n)),  \eqno(3b)
$$
under the condition that $dx^\mu b(x,n) = b(x,n)dx^\mu$. We have to assume
the rather unusual rule for $\chi_m$ that
$$
            a(x,m)\chi_m b(x,n) = a(x,m)b(x,m)\chi_m,  \eqno(4)
$$
which makes the product of matrix valued functions consistently calculable.
Eq.(4) is only an example to show how to change the discrete variable
of $b(x,n)$ when $\chi_m$ is shifted from left to right in an equation.
Such a change is necessary to assure the consistent product of matrices.
We call this rule the shifting rule which reflects the non-commutative
nature of the differential calculus just now treated.
According to this shifting rule, the third equation in Eq.(2) is rewritten as
$$
   d_{\chi_k} a(x,n) = [-a(x,n)M_{nk} + M_{nk}a(x,k)]\chi_k\,. \eqno(5)
$$
Eq.(5) and the shifting rule enable us to prove the Leibniz rule for $d_\chi$.
$$
d_\chi (a(x,n)b(x,n))=(d_\chi a(x,n))b(x,n)+a(x,n)(d_\chi b(x,n)), \eqno(3c)
$$
which together with Eq.(3b) serves to prove Eq.(3a).
\par
We investigate the nilpotency of ${\bf d}$ in the follows.
{}From the definition
of ${\bf d}$ in Eq.(2),
$$
 {\bf d}^2 a(x,n)=(d^2+ d\cdot d_\chi +d_\chi\cdot d +d_\chi^2)a(x,n). \eqno(6)
$$
$dx^\mu\wedge dx^\nu=-dx^\nu\wedge dx^\mu$
and $dx^\mu\wedge \chi_m=-\chi_m\wedge dx^\mu$
are reasonable in the ordinary differential calculus and yield
$$
      d^2a(x,n)=0, \hskip 1cm (d\cdot d_\chi +d_\chi\cdot d)a(x,n)=0\,.
\eqno(7)
$$
However, $\chi_m\wedge\chi_k$ is different basis from $-\chi_m\wedge\chi_k$
for $m\ne k$ due to the non-commutative property of geometry on the discrete
space. Therefore, ${d_\chi}^2a(x,n)=0$ can not simply result because of
the non-commutativity of $\chi_m\wedge\chi_k$.
However, the nilpotency is very important
   to guarantee the gauge invariant formulation
                     and so the proof of it is inevitable.
\par
 By remarking that $\chi_{m}$ appears in the form $M_{nm}\chi_m$ in Eq.(2),
we further define that
$$
 \eqalign{ d_{\chi_k} (M_{nm}\chi_m) &= M_{nm}M_{mk}\chi_{m}\wedge \chi_k, \cr
             d_{\chi_k}M_{nm}=M_{nk}&\chi_k M_{km}=M_{nk}M_{km}\chi_k\,.\cr
   } \eqno(8)
$$
Eq.(8) means $d_{\chi_k}\chi_m \ne 0$
which is also due to the non-commutativity
of the differential calculus on the discrete space.
We can easily calculate
$$ \eqalign{
  d_\chi^2 M_{nm} &= \sum_{l,k}d_{\chi_l}(M_{nk}{\chi_k}M_{km})\cr
    &=\sum_{k,l}\{(d_{\chi_l}(M_{nk}\chi_k))M_{km}-(M_{nk}\chi_k)
                          \wedge (d_{\chi_l}M_{km})\} \cr
    &=\sum_{k,l}\{(M_{nk}M_{kl}\chi_k\wedge \chi_l) M_{km}
                      -M_{nk}\chi_k\wedge (M_{kl}\chi_l M_{lm})\}
                           =0 , \cr}
                           \eqno(9)
$$
and
$$ \eqalign{
  d_\chi^2 a(x,n) &= \sum_{k,m}d_{\chi_k}\{-a(x,n)\,(M_{nm}\chi_m)
               +(M_{nm}\chi_m)\,a(x,n)\}\cr
  &=\sum_{k,m}[\,-(d_{\chi_k}a(x,n)\,)\wedge(M_{nm}\chi_m)
  -a(x,n)\,(d_{\chi_k}\,(\,M_{nm}\chi_m)\,)\, \cr
    &\hskip 1.5cm+
   (\,d_{\chi_k}\,(M_{nm}\chi_m\,)\,)\,a(x,n)
   -(M_{nm}\chi_m)\wedge(d_{\chi_k}a(x,n)\,)\,]=0,\cr
     } \eqno(10)
$$
where we also adopt such a reasonable rule that
$$
       d_{\chi_k}(M_{nm}\chi_m)a(x,m)=(\,d_{\chi_k}(\,M_{nm}\chi_m)\,)\,a(x,m)
          -(\,M_{nm}\chi_m\,)\wedge(\,d_{\chi_k}a(x,m)\,).
$$
In order to get the nilpotency of $d_\chi$ in Eqs.(9) and (10)
we have to carefully use the shifting rule.
With these equations, we can prove
$$
     {\bf d}^2 a(x,n)=0, \hskip 1cm {\bf d}^2 M_{nm}=0. \eqno(11)
$$
We can also prove the nilpotency ${\bf d}^2=0$ in the general case.
For example, we make the following manipulation for $a(x,n)M_{nm}$.
$$
\eqalign{
    {\bf d}^2 (a(x,n)M_{nm})&=d_\chi^2(a(x,n)M_{nm})\cr
           &=  {d_\chi}\,\{ \,(\,{d_\chi}a(x,n))M_{nm}
         +a(x,n)\,({d_\chi}M_{nm}\,)\,\} \cr
 &=\{ \,(\,{d_\chi}^2 a(x,n)\,)\,M_{nm}
    -(\,{d_\chi}a(x,n)\,)\,(\,{d_\chi}M_{nm}\,)\,\}\cr
   &    +\{\,({d_\chi} a(x,n)\,)\,(\,{d_\chi}M_{nm})
   +a(x,n)\,(\,{d_\chi}^2M_{nm}\,)\,\}=0. \cr
   }
$$
However, such a general case is not necessary in this paper and only  Eq.(11)
is quoted.

\par

Let us rewrite $A(x,n)$ in Eq.(1) by use of Eq.(2),
$$
     \eqalign{A(x,n) &\= \sum_{i}a_{i}^\dagger(x,n)
       [\partial_\mu a_{i}(x,n)dx^\mu +
                \sum_{m}(-a_i(x,n)M_{nm} + M_{nm}a_i(x,m))\chi_m] \cr
         &\= \sum_{i}a_{i}^\dagger(x,n)\partial_\mu a_{i}(x,n)dx^\mu \cr
        &\hskip 1.5cm + \sum_{m}\sum_{i}a_{i}^\dagger(x,n)
        \,(-a_i(x,n)M_{nm} + M_{nm}a_i(x,m))\chi_m.\cr}
  \eqno(12)
$$
{}From this equation we can define the gauge field $A_\mu (x,n)$ and Higgs
field
$\Phi_{nm}(x)$ as follows.
$$
  \eqalign{
    A_\mu(x,n) &\= \sum_{i}a_{i}^\dagger(x,n)\partial_\mu a_{i}(x,n), \cr
\Phi_{nm}(x) &\= \sum_{i}a_{i}^\dagger(x,n)\,(-a_i(x,n)M_{nm}
            + M_{nm}a_i(x,m)).\cr
}  \eqno(13)
$$
According to Chamseddine et al.[6] we here assume
$ \sum_{i}a_{i}^\dagger(x,n)a_{i}(x,n)\=1$
without loss of generality. With this assumption we can show that
$$
   A_\mu^\dagger(x,n) \= -A_\mu (x,n), \hskip 2cm
   \Phi_{nm}^\dagger (x) \= \Phi_{mn}(x),  \eqno(14)
$$
and perceive that the summation about $i$ in Eqs.(1) is necessary for
$A_\mu(x,n)$ not to become the pure gauge field.
With $A_\mu (x,n)$ and $\Phi_{nm}(x)$, the generalized gauge field $A(x,n)$ is
denoted by
$$
    A(x,n) \= A_\mu(x,n)dx^\mu + \sum_{m}\Phi_{nm}(x)\chi_m.  \eqno(15)
$$
\par
Generalized field strength $ F(x,n)$ is defined as
$$
     F(x,n) \= {\bf d}A(x,n) + A(x,n)\wedge A(x,n). \eqno(16)
$$
In order to get the explicit expression of $ F(x,n)$, we have to calculate
 ${\bf d}A(x,n)$ from Eq.(1).
$$
   \eqalign{
    {\bf d}A(x,n) &\= {\bf d}\,(\sum_{i}a^\dagger_{i}(x,n){\bf d}a_i(x,n))\cr
             & \= \sum_{i}{\bf d}a^\dagger_{i}(x,n) \wedge {\bf d}a_i(x,n),\cr}
   \eqno(17)
$$
because of the nilpotency ${\bf d}^2 a_i(x,n)= 0$ in Eq.(11).
Inserting Eq.(2) into Eq.(17) and using Eq.(13) we get
$$
   \eqalign{
    {\bf d}A(x,n) &\= \partial_\mu A_\nu(x,n) dx^\mu \wedge dx^\nu \cr
       &\hskip 0.5cm + \sum_m[\partial_\mu \Phi_{nm}(x)
          +  A_\mu(x,n)M_{nm} - M_{nm}A_\mu(x,m)]dx^\mu \wedge \chi_m \cr
      &\hskip 0.5cm +\sum_{k,m}[\Phi_{nk}(x)M_{km} + M_{nk}\Phi_{km}(x) \cr
  &\hskip 1cm +M_{nk}M_{km} - \sum_{i}a_{i}^\dagger(x,n)M_{nk}M_{km}a_{i}(x,m)]
        \chi_k \wedge \chi_m, \cr} \eqno(18)
$$
where we have used the relation $ \sum_{i}a_{i}^\dagger(x,n)a_{i}(x,n)\=1$.
The last term in the square bracket in front of $\chi_k \wedge \chi_m$
is
the auxiliary field  to be denoted by
$$ \eqalign{
  Y_{nk}(x)&= \sum_{i}a_{i}^\dagger(x,n)M_{nk}M_{kn}a_{i}(x,n),\cr
  Y_{nkm}(x)&= \sum_{i}a_{i}^\dagger(x,n)M_{nk}M_{km}a_{i}(x,m),
  \hskip 1cm {\rm for} \,\,n\ne m\cr
  }
 \eqno(19)
$$
which play an important role to ensure the gauge invariance of \ymh
as shown later.
On the other hand, $A(x,n) \wedge A(x,n)$ in Eqs.(12) can be calculated using
Eq.(15) as
$$
\eqalign{
    A(x,n) \wedge A(x,n) &\= A_\mu(x,n)A_\nu(x,n)dx^\mu \wedge dx^\nu \cr
          &\hskip 1cm
          + \sum_{m}(A_\mu(x,n)\Phi_{nm}(x) - \Phi_{nm}(x)A_\mu(x,m))dx^\mu
\wedge \chi_m \cr          &\hskip 1cm
          + \sum_{k,m}\Phi_{nk}(x)\Phi_{km}(x)\chi_k \wedge \chi_m. \cr}
          \eqno(20)
$$
After deriving Eqs.(18) and (20), we prescribe $M_{nn}=0$
and $\chi_n\wedge\chi_n=0$ which yields $\Phi_{nn}(x)=0$.

{}From Eqs.(18) and (20),
we obtain the generalized field strength $ F(x,n)$ to be
$$\eqalign{
 F(x,n) &= { 1 \over 2}F_{\mu\nu}(x,n)dx^\mu \wedge dx^\nu +
               \sum_{m}D_\mu\Phi_{nm}(x)dx^\mu \wedge \chi_m \cr
              & + \sum_{k\ne n}V_{nk}(x)\chi_k \wedge \chi_n \cr
     & + \sum_{k\ne n}\sum_{\{m\ne k,m\ne n\}}V_{nkm}(x)\chi_k \wedge \chi_m,
\cr}
                \eqno(21)
$$
where
$$
  \eqalign{
  F_{\mu\nu}(x,n)&=\partial_\mu A_\nu (x,n) - \partial_\nu A_\mu (x,n)
               + [A_\mu(x,n), A_\mu(x,n)],\cr
  D_\mu\Phi_{nm}(x)&=\partial_\mu \Phi_{nm}(x) -
  (\Phi_{nm}(x)+M_{nm})A_\mu(x,m)\cr
        & \hskip 1.5cm + A_\mu(x,n)(M_{nm} + \Phi_{nm}(x))\cr
  V_{nk}(x)&= (\Phi_{nk}(x) + M_{nk})(\Phi_{kn}(x) + M_{kn}) -
             Y_{nk}(x),\hskip 0.5cm {\rm for}\,\,k\ne n\cr
  V_{nkm}(x)&= (\Phi_{nk}(x) + M_{nk})(\Phi_{km}(x) + M_{km}) -
             Y_{nkm}(x). \hskip 0.5cm{\rm for}\,\,k\ne n, \quad m\ne n\cr}
     \eqno(22)
$$
If we define $H_{nm}(x) = \Phi_{nm}(x) + M_{nm},$ the last three
 equations above
are rewritten as
$$
  \eqalign{
  D_\mu\Phi_{nm}(x)&=D_\mu H_{nm}(x)=\partial_\mu H_{nm}(x)
       + A_\mu(x,n)H_{nm}(x) - H_{nm}(x)A_\mu(x,m)\cr
  V_{nk}(x)&= H_{nk}(x)H_{kn}(x) - Y_{nk}(x),\hskip 1cm
  {\rm for}\,\,k\ne n\cr
  V_{nkm}(x)&= H_{nk}(x)H_{km}(x) -   Y_{nkm}(x),\hskip 1cm
  {\rm for}\,\,k\ne n, \quad m\ne n\cr}
  \eqno(23)
$$
The function $H_{nm}(x)$ represents the unshifted
Higgs field,
whereas $\Phi_{nm}(x)$ denotes the shifted Higgs field
with vanishing vacuum expectation value.
Therefore,
the present formulation of the gauge theory in the
non-commutative geometry
would yield Yang-Mills lagrangian
with $M_{nm}$ which is the vacuum expectation values of $\Phi_{nm}(x)$.
This is the meaning that $M_{nm}$ determines the scale and pattern of the
spontaneous breakdown of gauge symmetry.
Owing to the discrete property of $Z_{\mathop{}_{N}}$,
there arises no additional
physical degrees of freedom as in the Kaluza-Klein theory.
\par
In order to construct gauge-invariant lagrangian
we have to investigate the gauge transformation property.
The gauge transformation is defined by
$$
 A^{g}(x,n)=g^{-1}(x,n)A(x,n)g(x,n)+g^{-1}(x,n){\bf d}g(x,n),
\eqno(24)
$$
where  $g\,(x,n)$ is the gauge function with respect to the corresponding
unitary group.
Eq.(19) defines the gauge transformation
law for $A_\mu(x,n)$ and $\Phi_{nm}(x)$,
$$
\eqalign{
   A_\mu^{g}(x,n) &\= g^{-1}(x,n)\partial_\mu g(x,n) +
                      g^{-1}(x,n)A_\mu(x,n)g(x,n) \cr
\Phi^g_{nm}(x)&\=g^{-1}(x,n)\Phi_{nm}(x)g(x,m)
+g^{-1}(x,n)(M_{nm}g(x,m)-g(x,n)M_{nm}).\cr}
\eqno(25)
$$
In terms of $H_{nm}(x)$,
we have
$$
H^{g}_{nm}(x)=g^{-1}(x,n)H_{nm}(x)g(x,m).\eqno(26)
$$
This looks like the usual gauge transformation law of Higgs
field.
In fact, Eq.(22) shows
that Higgs field receives from both gauge transformations defined on
the $n$-th and $m$-th sheets.
The gauge transformation of field strength is then subject to
$$
 F^{g}(x,n)=g^{-1}(x,n) F(x,n)g(x,n),\eqno(27)
$$
where we have to take the gauge transformation for $a_{i}(x,n)$ to be
$$
      a^{g}_{i}(x,n) \= a_{i}(x,n)g(x,n), \eqno(28)
$$
in order to get the consistent gauge transformation.
In components, Eq.(24) read
$$
\eqalign{
F_{\mu\nu}^g(x,n)&=g^{-1}(x,n)F_{\mu\nu}(x,n)g(x,n),\cr
D_\mu^g\Phi_{nm}(x)&=g^{-1}(x,n)D_\mu\Phi_{nm}(x)g(x,m),\cr
V^g_{nk}(x)&=g^{-1}(x,n)V_{nk}(x)g(x,n),\cr
V^g_{nkm}(x)&=g^{-1}(x,n)V_{nkm}(x)g(x,m).\cr}\eqno(29)
$$
In order for the third and fourth equations in Eq.(29) to be consistent,
Eq.(28) is inevitable.
With these gauge transformations,
the gauge invariant Yang-Mills-Higgs lagrangian can be constructed
in the next section. \par

\section{
{
\hskip 3.5cm
{\bf {{\iii}}. Gauge invariant lagrangian}
}}
In order to get the \ymh from the field strength $ F(x,n)$ which is the
differential two form, we have to decide the metric structure
of the discrete space $M_4 \times Z_N$.
The ordinary Minkowski
metric in $M_4$ is to be supplemented by
additional assumption that
the discrete directions are orthogonal to $M_4$
and the metric along the discrete directions are left arbitrary:
$$
\eqalign{
<dx^\mu, dx^\nu>&=g^{\mu\nu},\quad
g^{\mu\nu}={\rm diag}(1,-1,-1,-1),\cr
<\chi_n, dx^\mu>&=0,\cr
<\chi_n, \chi_m>&=-\alpha^2\delta_{nm}.\cr
}
\eqno(30)
$$
Then the inner products of two-forms
can be calculated by the formulae
$$
\eqalign{
<dx^\mu \wedge dx^\nu,
dx^\rho \wedge dx^\sigma>&=g^{\mu\rho}g^{\nu\sigma}-
g^{\mu\sigma}g^{\nu\rho},\cr
<dx^\mu \wedge \chi_n,
dx^\nu \wedge \chi_m>&=-\alpha^2g^{\mu\nu}\delta_{nm},\cr
<\chi_n \wedge \chi_k,\chi_m \wedge \chi_l>&
      =\alpha^4 \delta_{nm}\delta_{kl},\cr}
\eqno(31)
$$
while other inner products among the basis two-forms vanish.
It should be noticed that $\,\chi_m \wedge \chi_l\,$ is different basis
from $\,-\chi_l \wedge \chi_m\,$ for $m \ne l\,$ which reflects
 the non-commutative property of geometry on the discrete space.
Thus, the third equation in Eq.(31) is followed.
\par
According to these metric structures and Eq.(21) we can give
the formula for gauge-invariant \ymh
$$\eqalign{
{\cal L}_{{\mathop{}_{YMH}}}(x)&=-\sum_{n=1}^{\mathop{}_{N}}{1 \over g_{n}^2}
< F(x,n),  F(x,n)>\cr
=&-{\rm tr}\sum_{n=1}^{\mathop{}_{N}}{1\over 2g^2_n}
F_{\mu\nu}^{\dag}(x,n)F^{\mu\nu}(x,n)\cr
&+{\rm tr}\sum_{n=1}^{\mathop{}_{N}}\sum_{m\ne n}{\alpha^2\over g_{n}^2}
    (D_\mu\Phi_{nm}(x))^{\dag}D^\mu\Phi_{nm}(x)\cr
-{\rm tr}\sum_{n=1}^{\mathop{}_{N}}
{\alpha^4\over g_{n}^2}&\sum_{k\ne n}V_{nk}^{\dag}(x)
V_{nk}(x)
-{\rm tr}\sum_{n=1}^{\mathop{}_{N}}{\alpha^4\over g_{n}^2}
         \sum_{k\ne n}\sum_{m\ne n,m}V_{nkm}^{\dag}(x)V_{nkm}(x),\cr
}
\eqno(32)
$$
where $g_n$ is the gauge coupling constant
and
tr denotes the trace over internal symmetry matrices.
The third term is the potential term of Higgs particle and the fourth term
is the interaction term between Higgs particles.
Thus, the fourth term does not appear when $N=2$.
 $g_n$ is usually
taken to be common value for all $n$ in non-commutative geometry approach.
In our previous paper where two sheets ($y=\pm$) are prepared to reconstruct
\ws, we also took $g_y$ to be different values for $y=\pm$. We discussed there
possible quantum corrections which seems to relate
to the differences between $g_+$ and $g_-$.
\par
Let us turn to the fermion sector to construct the Dirac lagrangian
for leptons and quarks which is ordinarily obtained
by use of the Clifford algebra [4,5,6]. However, we want to get it in the same
way as \ymh in Eq.(28) is introduced
by using the inner product of the differential forms.
We achieved to give the Dirac lagrangian
in a way just mentioned now [8].
We shall show it in more suitable way to our approach in this paper.
\par
We define the covariant derivative acting on the spinor $\psi(x,n)$
by
$$
{\cal D}\psi(x,n)=({\bf d}+ A^f(x,n))\psi(x,n),
\eqno(33)
$$
which is called the generalized spinor one form.
Here $A^f(x,n)$
is the differential representation for the fermions
such that
$$
A^f(x,n)=A_\mu^f(x,n)dx^\mu+\sum_m \Phi^{f}_{nm}(x)\chi_m.
\eqno(34)
$$
It should be noticed that $A_\mu^f(x,n)$ is the differential representation
with respect to $\psi(x,n)$ and does not necessarily coincide with
$A_\mu(x,n)$ in the gauge boson sector. $\Phi_{nm}^{f}(x)$ is also the
corresponding expression to the representation of $\psi(x,n)$.
This will become important in $SU(5)$ model in sec.V.
We further define
$$
\eqalign{
d_\chi \psi(x,n)&=\sum_m d_{\chi_m} \psi(x,n)\cr
     &=\sum_m M_{nm}^{f}\chi_m\psi(x,n)=\sum_m M_{nm}^{f}\psi(x,m)\chi_m ,\cr}
\eqno(35)
$$
which leads to
$$
{\cal D}\psi(x,n)= (\partial_\mu + A^f_\mu(x,n)dx^\mu+\sum_m
H^f_{nm}(x)\chi_m)\psi(x,n),
\eqno(36)
$$
by use of $H^f_{nm}(x)=\Phi^f_{nm}(x)+M^f_{nm}$. $M_{nm}^f$ in Eq.(35)
is also the corresponding expression to $\Phi_{nm}^f(x)$.
\par
In the follows we investigate the gauge transformation property
of ${\cal D}\psi(x,n)$.
Let the gauge transformation of $\psi(x,n)$ to be
$$
   \psi^g(x,n) = (g^f(x,n))^{-1} \psi(x,n), \eqno(37)
$$
where $g^f(x,n)$ is the gauge transformation function corresponding
to the representation of $\psi(x,n)$.
Then, $A_\mu^f(x,n)$  is transformed as
$$
  {A_\mu^f}^g(x,n)=(g^f(x,n))^{-1}dg^f(x,n)
           +(g^f(x,n))^{-1}A^f_\mu(x,n)g^f(x,n),
\eqno(38a)
$$
and we choose $\psi(x,n)$ for the assignments
 of fermion in sections I\hskip -0.2mmV and V
    in order for $H^f_{nm}(x)$ to transform as
$$
 {H_{nm}^f}^g(x)\psi^g(x,m)=(g^f(x,n))^{-1}H_{nm}^f(x)\psi(x,m). \eqno(38b)
$$
{}From Eqs.(37) and (38a,b), we easily obtain the gauge covariance of ${\cal
D}\psi(x,n)$.
$$
     {\cal D}\psi^g(x,n)=(g^f(x,n))^{-1}{\cal D}\psi(x,n). \eqno(39)
$$
\par
Since ${\bf d}+ A^f(x,n)$
is Lorentz {\it invariant}\ ,
${\cal D}\psi(x,n)$
transforms as a spinor just like $\psi(x,n)$ does.
As the covariant derivative introduced above is the generalized
differential one form,
in order to get the Dirac lagrangian by use of the inner products
of differential forms
we introduce
the associated spinor one-form
by
$$
{\tilde {\cal D}}\psi(x,n)= \gamma_\mu \psi(x,n)dx^\mu
              -i{c_{ \mathop{}_{Y}}}\psi(x,n)\sum_m\chi_m,
\eqno(40)
$$
where ${c_{ \mathop{}_{Y}}}$
is a real,
dimensionless constant and invariant
against the gauge transformation.
${c_{ \mathop{}_{Y}}}$ will be shown to be related to the coupling constant
between Higgs particle and fermions and so the fermion mass.
Thus, its numerical value
is dependent on the fermions to be considered.
It is evident that ${\tilde {\cal D}}\psi(x,n)$ is gauge covariant
$$
      {\tilde {\cal D}}\psi^{g}(x,n) =(g^f(x,n))^{-1}{\tilde {\cal
D}}\psi(x,n).        \eqno(41)
$$
Dirac $\gamma$-matrices are taken to satisfy
$\gamma_\mu\gamma_\nu+\gamma_\nu\gamma_\mu=2g_{\mu\nu}$
with $\gamma_\mu^{\dag}=\gamma_0\gamma_\mu\gamma_0$.
The Lorentz transformation law for ${\tilde {\cal D}}(x,n)\psi(x,n)$
is the same
as that of $\psi(x,n)$ and so ${\cal D}\psi(x,n)$.\par
The Dirac lagrangian to be Lorentz and gauge invariant is computed by taking
the inner product
$$\eqalign{
{\cal L}_{ \mathop{}_{D}}(x,n)=&i<{\tilde {\cal D}}\psi(x,n),{\cal
D}\psi(x,n)>\cr
=&i\,{\rm
tr}\,[\,{\bar\psi}(x,n)\gamma^\mu(\partial_\mu+A_\mu^{f}(x,n))\psi(x,n)
+i{g_{ \mathop{}_{Y}}}{\bar\psi}(x,n)\sum_m H^f_{nm}(x)\psi(x,m)\,],\cr
}
\eqno(42)
$$
where ${-c_{ \mathop{}_{Y}}}\alpha^2$ is replaced simply by ${g_{
\mathop{}_{Y}}}$, and
we have defined the inner products for spinor one-forms, by
noting Eq.(30),
$$\eqalign{
<A(x,n)dx^\mu, B(x,m)dx^\nu>&={\rm tr}\,\bar{A}(x,n)B(x,m)g^{\mu\nu},\cr
<A(x,n)\chi_k, B(x,m)\chi_l>&
                  =-{\rm tr}\,\bar{A}(x,n)B(x,m)\alpha^2\delta_{kl},\cr
}
\eqno(43)
$$
with vanishing other inner products.
Here $\bar{A}(x,n)=A^{\dag}(x,n)\gamma^0$ dictates the usual Pauli adjoint
and Hermite conjugate with respect to the group representation. \par
Yukawa coupling of Higgs to fermions is given by
the last term of the second equation of Eq.(42):
$$
{\cal L}_{\mathop{}_{\rm Yukawa}}(x,n)=-{g_{ \mathop{}_{Y}}}\,
                  {\rm tr}\sum_m{\bar\psi}(x,n)H^f_{nm}(x)\psi(x,m),
\eqno(44)
$$
which contains Yukawa coupling constant ${g_{ \mathop{}_{Y}}}$ not present
in the bosonic sector.
We can also define
the Dirac lagrangian by ${\cal L}_{ \mathop{}_{D}}(x,n)=-i<{\cal D}\psi(x,n),
{\tilde {\cal D}}\psi(x,n)>$ which yields the same result as Eq.(42)
after the sum about  $n$.
The total Dirac lagrangian is the sum over $n=1,2,\cdots N$.
$$
{\cal L}_{\mathop{}_{D}}(x)
      =\sum_{n=1}^{\mathop{}_{N}}{\cal L}_{\mathop{}_{D}}(x,n),
\eqno(45)
$$
which is evidently ${\cal L}_{ \mathop{}_{D}}(x)^\dagger
={\cal L}_{ \mathop{}_{D}}(x)$ and so becomes
a real constant invariant against Lorentz and gauge transformation.
Eqs.(32) and (42) play an important role
to reconstruct the \ws and $SU(5)$ Gut in our formalism.
\section{
{
\hskip 3.1cm
{\bf I$\!$V.  $SU(3)_c \times SU(2) \times U(1)$ Standard model}
}}
Three sheets $(N=3)$ are necessary to reconstruct the standard model
corresponding to three gauge groups.
$n,m,k$ in Eqs.(32) and (42) then take the values from 1 to 3.
Let begin with the bosonic sector to write the explicit expression of Eq.(32).
$SU(2)\times U(1)$ part was discussed in great detail in Ref.[9].
We, thus, explain the model construction to stress how to incorporate
the strong interaction part $SU(3)_c$.
\par
\noindent
\centerline{ \bf{A. bosonic sector}}
\par
We assign $A(x,n), \Phi_{nm}(x),$ and $M_{nm} (n,m=1,2,3)$ as follows.
$$
\eqalign{
    A_\mu(x,1)&= -{i\over 2}\sum_{a=1}^8 \lambda^a G_\mu^a(x),\cr
    \Phi_{12}(x)&=\Phi_{13}(x)=\Phi_{21}(x)=\Phi_{31}(x)=0,\cr
    M_{12}=&M_{13}=M_{21}=M_{31}=0,\cr
    g(x,1)&=g_s(x),\hskip 1cm g_s(x)\in SU(3)_c,\cr}
    \eqno(46a)
$$
where $G_\mu^a(x)$ denotes the gluon field and the second and third equations
mean $SU(3)_c$ gauge symmetry not to break.
$$\eqalign{
A_\mu(x,2)&=-{i \over 2}\sum_{i=1}^3\tau^i A^i_\mu(x)-{i \over
2}a\tau^0B_\mu(x),\cr
\Phi_{23}(x)&=
\left(
\matrix{
\phi_+(x)\cr
\phi_0(x)\cr
}\right), \hskip 1cm
M_{23}=
\left(
\matrix{
0\cr
\mu\cr
}\right),\cr
 g(x,2)&=e^{-ia\alpha(x)}g(x),\;e^{-ia\alpha(x)}\in U(1),\;
g(x)\in SU(2),\cr
}
\eqno(46b)
$$
where $A_\mu^i(x)$ is $SU(2)$ gauge field and $B_\mu(x)$ is $U(1)$ gauge field.
$$\eqalign{
A_\mu(x,3)&=-{i \over 2}bB_\mu(x),\cr
\Phi_{32}(x)&=\Phi^{\dag}_{23}(x) \hskip 1cm
M_{32}=(0,\;\; \mu)=M^{\dag}_{23},\cr
g(x,3)&=e^{-ib\alpha(x)}\in U(1).\cr
}
\eqno(46c)
$$
where
$\mu$ in $M_{23}$ is a real, positive constant
which determines the scale of the symmetry breaking.
It should be noticed that there exist the free parameters $a,b$
in Eqs.(46a) and (46b). This choice of gauge fields for $n=2,3$
implies that $U(2)$ on the second sheet $n=2$ is not orthogonal
to $U(1)$ on the third sheet $n=3$,
 but the gauge group is only $SU(2) \times U(1)$.
\par
Above assignments yield
$$
\eqalign{
 F(x,1)&= -{i \over 4}\sum_a \lambda^a G_{\mu\nu}^a(x) dx^\mu \wedge dx^\nu,\cr
 F(x,2)&= {i\over 4}[\,- \sum_i\tau^i F_{\mu\nu}^i(x)
          -a\tau^0 B_{\mu\nu}(x)\, ]dx^\mu \wedge dx^\nu\cr
               + & D_\mu H(x)dx^\mu \wedge \chi_3
         +[\,H(x)H^\dagger(x)-Y_{23}(x)\,]\chi_2\wedge\chi_3,\cr
  F(x,3)&= -b{i\over 4}B_{\mu\nu}(x)dx^\mu \wedge dx^\nu \cr
            &+(\,D_\mu H(x)\,)^\dagger dx^\mu \wedge \chi_2
      + [\,H^\dagger(x)H(x)-Y_{32}(x)\,]\chi_3\wedge\chi_2,\cr
   } \eqno(47)
$$
where
$$
\eqalign{
      G_{\mu\nu}^a(x)=&\partial_\mu G_\nu^a(x)-\partial_\nu G_\mu^a(x)
         +f^{abc}G_\mu^b(x)G_\nu^c(x),  \cr
      F_{\mu\nu}^i(x)=&\partial_\mu A_\nu^i(x)-\partial_\nu A_\mu^i(x)
         +\epsilon^{ijk}A_\mu^j(x)A_\nu^k(x),  \cr
      B_{\mu\nu}(x)=&\partial_\mu B_\nu(x)-\partial_\nu B_\mu(x),\cr
      H(x)=&\Phi_{23}(x)+M_{23}.\cr
      D^\mu H(x)=&[\,\partial_\mu-{i\over 2}\,(\sum_i\tau^iA_\mu^i(x)
          +(a-b)\,\tau^0\,B_\mu(x)\,)\,]\,H(x). \cr
      } \eqno(48)
$$
A straightforward calculation yields,
$$\eqalign{
&{\cal L}_{{\mathop{}_{YMH}}}=-\sum_{n=1}^3{1\over g_n^2}<F(x,n), F(x,n)>\cr
=&-{1\over 4g_1^2}\sum_a G_{\mu\nu}^a(x)G^{a\mu\nu}(x)\cr
&-{1\over 4g_2^2}
\sum_i F^i_{\mu\nu}(x)\cdot F^{i\mu\nu}(x)
-{1\over4}
({a^2\over g_2^2}+{b^2\over 2g_3^2})B_{\mu\nu}(x)B^{\mu\nu}(x))\cr
&+({1\over g_2^2}+{1\over g_3^2})\alpha^2(D_\mu H(x))^{\dag}D^\mu H(x)\cr
&-{\alpha^4\over g^2_2}\,{\rm tr}\,(H(x)H^{\dag}(x)-Y_{23}(x))^2
-{\alpha^4\over g_3^2}(H^{\dag}(x)H(x)-Y_{32}(x))^2.\cr
}
\eqno(49)
$$
Let us investigate the auxiliary fields $Y_{23}(x)$ and $Y_{32}(x)$
in Eq.(49) from Eq.(19).
$$
  \eqalign{
   Y_{23}(x) &= \sum_{i}a_i^\dagger(x,2)M_{23}M_{32}a_i(x,2) \cr
          &= \sum_{i}a_i^\dagger(x,2)
          \left(\matrix{0 & 0 \cr
                        0 & \mu^2 \cr} \right) a_i(x,2),\cr} \eqno(50)
$$
which is the independent field. Therefore, we discard the potential term
involving $Y_{23}(x)$ in Eqs.(49) using the equation of motion.
On the other hand,
$$
  \eqalign{
   Y_{32}(x) &= \sum_{i}a_i^\dagger(x,3)M_{32}M_{23}a_i(x,3) \cr
          &= \sum_{i}a_i^\dagger(x,3) \mu^2 a_i(x,3)= \mu^2,\cr}\eqno(51)
$$
is dependent so that the potential term containing $Y_{32}(x)$ remains.
\par
After rescaling
$$
\eqalign{
G_\mu(x) &\rightarrow g_3 G_\mu(x),\cr
A_\mu^i(x)  &\rightarrow g_2 A_\mu^i(x),\cr
B_\mu(x) &\rightarrow {g_2g_3\over
      \sqrt{a^2g_3^2+{\dis{b^2\over 2}}g_2^2}}B_\mu(x),\cr
H(x) &\rightarrow{g_2g_3\over \sqrt{g_2^2+g_3^2}\alpha}H(x)
         =g_{\mathop{}_{H}}H(x)\cr
}
$$
we find that
$$\eqalign{
{\cal L}_{{\mathop{}_{YMH}}}=&-{1\over 4}\sum_a G_{\mu\nu}^a(x)\cdot
G^{a\mu\nu}(x)
  -{1\over 4}\sum_i F_{\mu\nu}^i(x)\cdot F^{i\mu\nu}(x)
  -{1\over 4} B_{\mu\nu}(x)\cdot B^{\mu\nu}(x) \cr
&+(D_\mu H(x))^{\dag}(D^\mu H(x))
-
\lambda(\,H^{\dag}(x)\,H(x)
-\mu'^2
)^2
,\cr
}
\eqno(52)
$$
where
$$
\eqalign{
G_{\mu\nu}^a (x) &= \partial_\mu G_\nu^a(x)-\partial_\nu G_\mu^a(x)
       +g_1f^{abc}G_\mu^b G_(x)\nu^c(x) \cr
F_{\mu\nu}^i (x) &= \partial_\mu A_\nu^i(x)-\partial_\nu A_\mu^i(x)
       +g_2\epsilon^{ijk}A_\mu^j A_(x)\nu^j(x) \cr
D_\mu H(x)&=[\partial_\mu-{i \over 2}(g_2\sum_i \tau^i\cdot  A^i_\mu(x)
+g'\tau_0B_\mu(x))]H(x),\cr}
\eqno(53)
$$
with $\dis{g^{'}={(a-b)g_2g_3\over \sqrt{a^2g_3^2+{b^2\over 2}g_2^2}}}\,$,
$\dis{\lambda={g_2^{4}g_3^{2} \over (g_2^2+g_3^2)^2}\,}$
and $\dis{{\sqrt{g_2^2+g_3^2}\alpha\mu \over g_2g_3}}$ is simply replaced
by $\mu'\,$.
This is \ymh
with $SU(3)_c\times SU(2)_L \times U(1)$ spontaneously
broken to $SU(3)_c\times U(1)_{\rm em}$.
The gauge coupling constant of $U(1)$ is related to that of $SU(2)_L$
through the parameters $a, b, {\delta={g_2\over g_3}}$.
Putting $W_\mu={1\over \sqrt{2}}(A_\mu^1-iA_\mu^2),
\, W_\mu^{\dag}={1\over \sqrt{2}}(A_\mu^1+iA_\mu^2)$
and
 $Z_\mu={1\over {\sqrt{g^2+g^{'2}}}}(-gA_\mu^3+g'B_\mu)$,
we obtain the gauge boson mass terms
$$
{\cal L}_{{\rm gauge}\;{\rm boson}\;{\rm mass}}
=m_{\mathop{}_{W}}^2W_\mu^{\dag}W^\mu+{1\over 2}m_{ \mathop{}_{Z}}^2Z_\mu
Z^\mu,
\eqno(54)
$$
where
$$
 \eqalign{
 m_{ \mathop{}_{W}}&=\sqrt{{{1 + \delta^2} \over 2}}\alpha\mu,\cr
 m_{ \mathop{}_{Z}}&=\sqrt{{1+\delta^2 \over 2}(
 1+{2(a^-b)^2\over {2a^2+\delta^{2}b^2}})}\alpha\mu.\cr}\eqno(55)
$$
Weinberg angle is
determined by
two parameters; $r={b\over a}$ and $\delta$:
$$
\sin^2{\theta_{ \mathop{}_{W}}}=1-{m_{ \mathop{}_{W}}^2\over
m_{ \mathop{}_{Z}}^2}={2(1-r)^2\over
{4-4r+(2+\delta^2)r^2}},
\eqno(56)
$$
which becomes $\;\dis{{1 \over 2+2\delta^2}}\;$ for
$r=2$ ($b=2a$) derived from the so called super-traceless condition,
${\rm tr}A_\mu(x,2)={\rm tr}A_\mu(x,3)$ [4,5].
On the other hand,
the Higgs mass is given from Eq.(52) by
$$
   m_{ \mathop{}_{H}} = { 2\delta \over \sqrt{1+\delta^2}}\alpha\mu, \eqno(57)
$$
which is related to the $W$ gauge boson mass
via
$$
m_{ \mathop{}_{H}}={2\sqrt{2}\delta \over 1+\delta^2}m_{ \mathop{}_{W}}.
\eqno(58)
$$
Eqs.(58) leads to the inequality $m_{ \mathop{}_{H}} \leq
{\sqrt 2}m_{ \mathop{}_{W}}$.
If we set $\delta=1$, $m_{ \mathop{}_{H}}={\sqrt 2}m_{ \mathop{}_{W}}$,
which is typical [4,5,]
in the non-commutative geometry and the maximal value of
Higgs boson mass $m_{ \mathop{}_{H}}$ in this model. These results are already
deduced in Ref.[9] in the slightly different formalism where only two sheets
are incorporated and
so gluon fields are not taken into account. Model \ii\  in Ref.[9]
is more interesting where the ratio $r={b\over a}$ is decided to be 2
from the physical
requirement that the electromagnetic interaction is not broken spontaneously.
This leads us to the interesting relation
$$
         m_{ \mathop{}_{H}}=2\sqrt{2}m_{ \mathop{}_{W}}
         \sin \theta_{ \mathop{}_{W}}, \eqno(59)
$$
though we do not discuss the model \ii\  in Ref.[9] any more.
\par
\noindent
\centerline{\bf B.\  fermionic sector}
\par

Let us
reconstruct the fermionic sector of W-S theory using the formalism presented in
section \iii. We first consider the leptonic sector in the first generation.
Since leptons do not interact with gluons through the strong interaction,
we assign $\psi(x,1)=0$.
 In conformity with the chiral nature of leptons
we assign $SU(2)$ doublets
with left handed chirality on the second sheet $n=2$, and
$SU(2)$ singlets with right handed chirality on the third sheet $n=3$.
Consequently, in this context
$A^f_\mu(x,2)$ is given by
$A_\mu(x,2)$ in Eq.(46b) with $a\rightarrow a_{\mathop{}_{L}}$
and $A^f_\mu(x,3)$ by $A_\mu(x,3)$ in Eqs.(46c)
with $b\rightarrow a_{\mathop{}_{R}}$,
where $a_{\mathop{}_{L,R}}$ are related to hypercharges
of the leptons. In order that the $U(1)$ invariance is consistently realized
in Eqs.(44), $b-a=a_{\mathop{}_{R}}-a_{\mathop{}_{L}}$ is necessary.
In view of Eqs.(53) in which gauge fields and Higgs particle are scaled out
we may set $a_{\mathop{}_{L,R}}=Y$ where $Y$ is the hypercharge of lepton.
With this in mind we make the following assignments for the leptons
in first generation.
\par
$$
\psi(x,2)=l_{\mathop{}_{L}}(x)=\left(
          \matrix{
          \nu_{\mathop{}_{L}}(x)\cr
          e_{\mathop{}_{L}}(x)\cr
          }\right),\quad a_{\mathop{}_{L}}=-1.
\eqno(60a)
$$
$$
\psi(x,3)=l_{\mathop{}_{R}}(x)=e_{\mathop{}_{R}}(x),\quad a_{\mathop{}_{R}}=-2,
\eqno(60b)
$$
where $\nu$ stands for $\nu_e$
and
the left and right-handed
fields are defined by
$\psi_{\mathop{}_{L}}={{1-\gamma_5}\over 2}\psi$ and
$\psi_{\mathop{}_{R}}={{1+\gamma_5}\over 2}\psi$, respectively,
with $\gamma_5=i\gamma^0\gamma^1\gamma^2\gamma^3$.
To be more precise the left-handed leptons form a doublet with
hypercharge $Y=a_{\mathop{}_{L}}=-1$
and
the right-handed
electron a singlet with
hypercharge $Y=a_{\mathop{}_{R}}=-2$ as usual.
Gauge invariance of Yukawa coupling (44)
is satisfied by $a_{\mathop{}_{L}}-a_{\mathop{}_{R}}=a-b=1$.
With these assignments, Dirac lagrangian for leptonic sector can be written as
$$
\eqalign{
        {\cal L}_{\mathop{}_{D}}(x)=&i\,[\,\,{\bar l}_{\mathop{}_{L}}(x)
            \, \gamma^\mu\,(\partial_\mu-{i\over 2}g_2\sum_i \tau^iA_\mu^i(x)
                       +{i\over 2}g'\tau^0B_\mu(x)\,)\,l_{\mathop{}_{L}}(x)\cr
                &\hskip 3.5cm +{\bar l}_{\mathop{}_{R}}(x)\,
        \gamma^\mu\,(\partial_\mu+ig' B_\mu(x)\,)\,l_{\mathop{}_{R}}(x)\,\,]\cr
&-g_{\mathop{}_{Y}}[\,\,{\bar l}_{\mathop{}_{L}}(x)\,H(x)\,l_{\mathop{}_{R}}(x)
    +{\bar l}_{\mathop{}_{R}}(x)\,H^\dagger(x)\,l_{\mathop{}_{L}}(x)\,\,]\,.\cr
}
 \eqno(61)
$$
The mass term becomes simply
$$\eqalign{
{\cal L}_{{\rm lepton}\;{\rm mass}}=&-g_{\mathop{}_{Y}}\,
   [\,({\bar\nu}_{\mathop{}_{L}}, {\bar e}_{\mathop{}_{L}})
     M_{23}e_{\mathop{}_{R}}+{\bar e}_{\mathop{}_{R}}M_{32}\left(
                              \matrix{
                              \nu_{\mathop{}_{L}}\cr
                              e_{\mathop{}_{L}}\cr
                              }\right)]\cr
&=-m_e({\bar e}_{\mathop{}_{L}}e_{\mathop{}_{R}}
      +{\bar e}_{\mathop{}_{R}}e_{\mathop{}_{L}})=-m_e\bar ee\,,\cr
}
\eqno(62)
$$
where $m_e=g_{\mathop{}_{Y}}\mu$ as usual.
The electron becomes massive as expected whereas
the neutrino remains massless by the assignment Eqs.(60a,b).
To include three generations is very easy for leptonic sector
and so we skip it.
\par

Let us consider quarks in three generations.
In contrast to the leptonic sector we have to take into account the strong
interactions between quarks and gluons  and also give mass to the up quark.
Adding to these we must consider quark mixing between three generations.
With expressions of $q_{\mathop{}_{F}}^{'}=(d^{'}, s^{'}, b^{'})$
and $q_{\mathop{}_{f}}=(d, s, b)$, the quark mixing is expressed as
$$
          q_{\mathop{}_{F}}'=\sum_{f=1}^3 U_{\mathop{}_{Ff}}q_f, \eqno(63)
$$
where $U$ is Kobayashi-Maskawa mixing matrix.
Two kinds of assignments are necessary for quark sector to give masses
to up and down quarks.
At first, we make the following assignment to give mass to down quark.
\par
$$
\eqalign{
     \psi(x,1)&=0, \cr
    \psi(x,2)&=a_1\,q_{\mathop{}_{L}}=a_1\left(
          \matrix{
          u_{\mathop{}_{L}}(x)\cr
          d'_{\mathop{}_{L}}(x)\cr
          }\right),\quad a_{\mathop{}_{L}}={1 \over 3},\cr
     \psi(x,3)&= d'_{\mathop{}_{R}}(x),
             \quad a_{\mathop{}_{R}}=-{2 \over 3},\cr}
\eqno(64)
$$
where we abbreviate subscripts about triplet representation of $SU(3)_c$ and
$a_1$ exists for normalization of quark fields in the final expression.
We have to carefully investigate the differential representation $A^f(x,n)$
in eq.(34). $\psi(x,2)$ is triplet for $SU(3)_c$, doublet for $SU(2)$
and of course singlet for $U(1)$
and transformed according to $g(x,1)\otimes g^f(x,2)\,$,
where  $g^f(x,2)$ is given as $e^{-ia_{\mathop{}_{L}}\alpha(x)}g(x)$.
$\psi(x,3)$ is triplet for
$SU(3)_c$ and singlet for $SU(2)\times U(1)$ and transformed according to
$g(x,1)\otimes g^f(x,3)\,$,
where $g^f(x,3)$ is $e^{-ia_{\mathop{}_{R}}\alpha(x)}$.
As a result $A^f(x,n)$ is determined to be
$$
\eqalign{
    A^f(x,2)&=(-{i\over 2}g_1\sum_{a=1}^8 \lambda^aG_\mu^a(x)
    -{i\over 2}g_2\sum_{i=1}^3 \tau^iA_\mu^i(x)
                 -{i\over 6}g'\tau^0 B_\mu(x))dx^\mu
                 +g_{\mathop{}_{H}}H(x)\chi_3
                  ,\cr
    A^f(x,3)&=\,(-{i\over 2}g_1\sum_{a=1}^8 \lambda^aG_\mu^a(x)
              +{i\over 3}g'B_\mu(x)\,)\,dx^\mu
               +g_{\mathop{}_{H}} H^\dagger(x)\chi_2\,. \cr
    }\eqno(65)
$$
The associated spinor one form is taken to be
$$
{\tilde {\cal D}}\psi(x,n)= \gamma_\mu \psi(x,n)dx^\mu
              -ic_d\psi(x,n)\sum_m\chi_m\,.
\eqno(66)
$$
Adding to this assignment we make one more assignment
to give mass to up-quark.
$$
\eqalign{
     \psi'(x,1)&=0, \cr
    \psi'(x,2)&=a_2\left(
          \matrix{
          d_{\mathop{}_{R}}^c(x)\cr
         - {u'_{\mathop{}_{R}}}^c(x)\cr
          }\right),
               \quad a_{\mathop{}_{R}}={-1 \over 3},\cr
     \psi'(x,3)&=u'^c_{\mathop{}_{L}}(x),
     \quad a_{\mathop{}_{L}}=-{4 \over 3},\cr}
\eqno(67)
$$
where superscripts $c$ on quark fields represent the charge conjugation and
$a_1^2+a_2^2=1$ is satisfied for normalization of quark field.
$u'$ denotes the up-quark mixed by  $U^\dagger$ matrix. It should be noted
that $\psi'(x,2)$ is the right handed spinor and $\psi'(x,3)$, left handed.
$\psi'(x,2)$ is ${\underline {3}}^\ast$  representation
for $SU(3)_c$, doublet for $SU(2)$
and of course singlet for $U(1)$, and then transforms according to
$g(x,1)^\ast\otimes g^f(x,2)\,$,
where  $g^f(x,2)$ is given as $e^{-ia_{\mathop{}_{R}}\alpha(x)}g(x)$.
$\psi'(x,3)$ is ${\underline {3}}^\ast$  representation for
$SU(3)_c$ and singlet for $SU(2)\times U(1)$, and trasforms according to
 $g(x,1)^\ast\otimes g^f(x,3)$,
 where $g^f(x,3)$ is $e^{-ia_{\mathop{}_{L}}\alpha(x)}$.
For this assignment, $A^{f'}(x,n)$ are
$$
\eqalign{
    A^{f'}(x,2)&=({i\over 2}g_1\sum_{a=1}^{ 8} \lambda^{\ast a}G_\mu^a(x)
    -{i\over 2}g_2\sum_{i=1}^3 \tau^iA_\mu^i(x)
                 \! +\!{i\over 6}g'\tau^0 B_\mu(x)\,)dx^\mu
                 \!+\!g_{\mathop{}_{H}}H(x)\chi_3 ,\cr
    A^{f'}(x,3)&=(\,{i\over 2}g_1\sum_{a=1}^{8} \lambda^{\ast a}G_\mu^a(x)
           +{2i\over 3}g'B_\mu(x)\,)dx^\mu
            +g_{\mathop{}_{H}}H^\dagger(x)\chi_2. \cr
    }\eqno(68)
$$
The associated spinor one form is taken to be
$$
{\tilde {\cal D}}\psi'(x,n)= \gamma_\mu \psi'(x,n)dx^\mu
              -ic_u\psi'(x,n)\sum_{m=1}^3\chi_m,
\eqno(69)
$$
With these considerations, we can obtain the final expression as follows.
$$
        {\cal L}^q = {\cal L}^{q}_{kin.}+{\cal L}^{q}_{\mathop{}_{Yukawa}},
\eqno(70a)
$$
where
$$
\eqalign{
{\cal L}^{q}_{kin.}&=i{\bar{q}}_{\mathop{}_{L}}\,\gamma^\mu\,
               (\,\partial_\mu  -{i\over 2}g_1\sum_{a=1}^8 \lambda^a\,G_\mu^a
    -{i\over 2}g_2\sum_{i=1}^3 \tau^iA_\mu^i
                 -{i\over 6}\,g'\,\tau^0\, B_\mu\,)\,q_{\mathop{}_{L}}\cr
            &+i{\bar u'_{\mathop{}_{R}}}\,\gamma^\mu\,
            (\,\partial_\mu-{i\over 2}g_1\sum_{a=1}^8 \lambda^a\,G_\mu^a
               -i{2\over3}g'B_\mu)\,u'_{\mathop{}_{R}}\cr
         &+ i{\bar d'_{\mathop{}_{R}}}\,\gamma^\mu\,
         (\,\partial_\mu-{i\over 2}g_1\sum_{a=1}^8 \lambda^a\,G_\mu^a
         +i{1\over3}g'B_\mu\,)\,d'_{\mathop{}_{R}}\,,\cr}
\eqno(70b)
$$
and
$$
{\cal L}^{q}_{\mathop{}_{Yukawa}}
       =-g_{\mathop{}_{Y}}\,(\,{\bar q}_{\mathop{}_{L}}\,H\,d'_{\mathop{}_{R}}
              + {\bar d'}_{\mathop{}_{R}}\,H\,q_{\mathop{}_{L}}\,)
  -g'_{\mathop{}_{Y}}\,(\,{\bar q'}_{\mathop{}_{L}}\,
              {\tilde H}\,u'_{\mathop{}_{R}}
       + {\bar u'}_{\mathop{}_{R}}\,{\tilde H}\,q'_{\mathop{}_{L}}\,)\,,
\eqno(70c)
$$
with $g_{\mathop{}_{Y}}$ and $g'_{\mathop{}_{Y}}$ appropriately adjusted
coupling constants.
$q'_{\mathop{}_{L}}$ and ${\tilde H}$ in Eq.(70c) are given by
$$
           q'_{\mathop{}_{L}} =\left(\matrix{
                           u'_{\mathop{}_{L}} \cr
                           d_{\mathop{}_{L}} \cr}
                           \right), \hskip 1cm
            {\tilde H}=i\tau^2H^\ast.
$$
Eq.(70a) is Dirac lagrangian for first generation and we should add the second
and third generations to obtained the total Dirac lagrangian
though $q'_{\mathop{}_{L}}$ is already set to be $q_{\mathop{}_{L}}$
in Eq.(70b).

\par

\section{
{
\hskip 4.35cm
{\bf V.  $SU(5)$  grand unified theory}
}}
$SU(5)$ Gut is very attractive to account for many physically
interesting points that the standard model could not touch. For example,
it yields the charge quantization of quarks and predicts the numerical value
of Weinberg angle. In addition, it is important as a first model to unify
the strong, weak and electromagnetic interactions and it has long been
believed to be a true theory in the elementary particle physics. However,
it has been nowadays ruled out as a true theory because of no detection
of proton decay and accurate experiments of Weinberg angle by LEP.
However, it is even now meaningful to reconstruct $SU(5)$ Gut
in our formalism based on non-commutative geometry
since it is a pattern of grand
unified theory, and it is also necessary to construct the more realistic
model like $SO(10)$ unification model. \cham [6] succeeded to treat
$SU(5)$ model in their formalism. Our previous paper [9] is also
successful in reconstructing the bosonic sector of $SU(5)$ model but not
the fermionic sector. The purpose of this section is to present the formulation
of the fermionic sector although the bosonic sector is also explained to make
 this paper self-contained. Let begin with the bosonic sector.\par
\noindent
\centerline{\bf A.\ bosonic sector}
\par
According to \cham [6] we prepare
three copies of four dimensional Minkowski space $(\,N=3 )$. The symmetry
breakdowns in $SU(5)$ Gut occur at two scales. At the first stage $SU(5)$
is spontaneously broken to $SU(3)_c \times SU(2) \times U(1)$ with the grand
unified scale $M$
and again at the second stage broken to $SU(3)_c\times U(1)_{em}$
with the electro-weak scale $\mu$. The first stage of breakdown occurs through
the adjoint $\underline{24}$ Higgs representation
and the second stage through the fundamental $\underline{5}$
Higgs representation of $SU(5)$.
To nicely incorporate the first symmetry breakdown, we have to
identify two copies of three, which yields
$$
                      a_i(x,1)=a_i(x,2).             \eqno(71)
$$
{}From Eq.(13), these are taken to be $5\times 5$ matrices to give the $SU(5)$
gauge field and the adjoint Higgs field. To obtain the fundamental Higgs field
, $a_i(x,3)$ are taken to be $1\times 1$ matrices and, to avoid an extra
 $U(1)$ gauge field, these have to be real constants. Corresponding to these
 choices we have to decide the symmetry breaking patterns which are due to
 $M_{nk}$ in Eq.(5). $M_{12}=M_{21}$ cause the first stage breakdown and are
 given as
\font\fontB=cmr10 scaled \magstep3
\def\bigzerol{\smash{\lower-2ex\hbox{\fontB{0}}}}
\def\bigzerou{%
         \smash{\lower2ex\hbox{\fontB{0}}}}
$$
 M_{12} = M_{21} = M\left (\hskip -3mm
               \matrix{
              2 &   &   &    & \hskip -0.5cm \bigzerou \cr
                & \hskip -2mm 2 &   &    &           \cr
                &   & \hskip 1mm 2 &    &           \cr
                &   &   & -3 &           \cr
      \hskip 0.5cm \bigzerol &   &   &    & -3        \cr}
                  \right ) = \Sigma_0 \eqno(72)
$$
$M_{13}=M_{23}$ are responsible for the second stage breakdown
and are written as
$$
    M_{13}=M_{23}=M_{31}^\dagger=M_{32}^\dagger=\mu \left(
                \matrix{
                 0 \cr
                 0\cr
                 0\cr
                 0\cr
                 1\cr}
                 \right) =H_0 .   \eqno(73)
$$
With these assignment, the generalized gauge fields are expressed
by use of the gauge fields and Higgs fields as
$$
\eqalign{
      A(x,1)&=-{i\over 2}\sum_{a=1}^{24} T^a A_\mu^a(x)dx^\mu
           +\Sigma(x)\chi_2 + H(x)\chi_3, \cr
      A(x,2)&=-{i\over 2}\sum_{a=1}^{24} T^a A_\mu^a(x)dx^\mu
           +\Sigma(x)\chi_1 + H(x)\chi_3, \cr
      A(x,3)&=H^\dagger(x)\chi_1+H^\dagger(x)\chi_2, \cr
  }  \eqno(74)
$$
where $T^a$ is the generator of $SU(5)$ group, $A_\mu^a(x)$ is the $SU(5)$
 gauge field, $\Sigma(x)$ is the adjoint Higgs field
 and $H(x)$ is the fundamental Higgs field. It should be noticed that
 the gauge  field on the third sheet $A_\mu(x,3)$ vanishes
 because of the $x$ independent
 $a_i(x,3)$. \par
Above assignments deduce the following \ymh.
$$
\eqalign{
{\cal L}_{\mathop{}_{YMH}}(x)&=-{\rm tr}{1\over g^2}
                         F_{\mu\nu}(x)F^{\mu\nu}(x) \cr
   +{\rm tr}{2\alpha^2\over g^2}&(D_\mu\Sigma(x)\,)^\dagger(\,D_\mu\Sigma(x)\,)
   +   2\alpha^2({1\over g^2}+{1\over g'^2})
            (D_\mu H(x)\,)^\dagger(\,D_\mu H(x)\,) \cr
      & -V_1(\Sigma)-V_2(H)-V_3(\Sigma,H), \cr
 } \eqno(75)
$$
where
$$
\eqalign{
      D_\mu\Sigma(x)&=\partial_\mu \Sigma(x) +\,[\,A_\mu(x),\,
                 \Sigma(x)+\Sigma_0\,]\,,\cr
     D_\mu H(x) &=\partial_\mu H(x) + A_\mu(x)\,(\,H(x)+H_0\,)\,,\cr} \eqno(76)
$$
with $A_\mu(x)=-\dis{{i\over 2}\sum_{a=1}^{24}T^aA_\mu^a(x)}$ and
$$
\eqalign{
   V_1(\Sigma)&={\rm tr}{2\alpha^4\over g^2}\{\,(\,\Sigma(x)+\Sigma_0\,)^2 -
             Y_{12}(x)\,\}^2, \cr
 V_2(H)&={\rm tr}{2\alpha^4\over g^2}|\,(\,H(x)+H_0\,)\,(\,H(x)+H_0\,)^\dagger
-
             Y_{13}(x)\,|^2 \cr
    &+     {2\alpha^4\over g'^2}|\,(\,H(x)+H_0\,)^\dagger(\,H(x)+H_0\,) -
             Y_{31}(x)\,|^2, \cr
  V_3(\Sigma,H)&={\rm tr}{2\alpha^4\over g^2}|\,(\,\Sigma(x)+\Sigma_0\,)\,
                (\,H(x)+H_0\,)\,-Y_{123}(x)\,|^2 \cr
                &+{\rm tr}{2\alpha^4\over g^2}|\,(\,H(x)+H_0\,)\,
                (\,H(x)+H_0\,)^\dagger\,-Y_{132}(x)\,|^2 \cr
       &+{\rm tr}{2\alpha^4\over g'^2}|\,(\,H(x)+H_0\,)^\dagger\,
                (\,\Sigma(x)+\Sigma_0\,)\,-Y_{312}(x)\,|^2. \cr
} \eqno(77)
$$
In order to obtain the physical \ymh  the auxiliary fields $Y_{nk}(x)$, and
$Y_{nmk}(x)$ have to be evaluated to investigate which fields are independent.
$$
\eqalign{
    Y_{12}(x)&=\sum_i a_i^\dagger(x,1)M_{12}M_{21}a_i(x,1)\cr
       &={1\over 5}{\rm Tr} \Sigma_0^2-M(\,\Sigma(x)+\Sigma_0\,),\cr
       } \eqno(78)
$$
where use has been made Eqs.(13), (19) and (72) and so this field is dependent
left in the potential.
$$
\eqalign{
    Y_{13}(x)&=\sum_i a_i^\dagger(x,1)M_{13}M_{31}a_i(x,1)\cr
       &=\sum_i a_i^\dagger(x,1)H_0 H_0^\dagger a_i(x,1)\,,\cr}\eqno(79)
$$
which is evidently independent field and the potential term containing this
field is discarded owing to the equation of motion.
$$
\eqalign{
    Y_{31}(x)&=\sum_i a_i^\dagger(x,3)M_{31}M_{13}a_i(x,3)\cr
       &=\sum_i a_i^\dagger(x,1)H_0^\dagger H_0a_i(x,1)\,=\mu^2,\cr}\eqno(80)
$$
which is simply a number.
$$
\eqalign{
    Y_{123}(x)&=Y_{312}(x)^\dagger=
        \sum_i a_i^\dagger(x,1)M_{12}M_{23}a_i(x,3)\cr
       &=-3M\sum_i a_i^\dagger(x,1)H_0a_i(x,3)\,
          =-3M\,(\,H(x)+H_0\,),\cr} \eqno(81)
$$
where use has been made Eqs.(13), (19), (72) and (73) and so this field is
dependent left in the potential. \par
\noindent
As $Y_{132}(x)=Y_{13}(x)$, $Y_{132}(x)$ is independent field so that this term
is eliminated. With these considerations we can obtain the final expression
for the potential term in Eq.(75).
$$
\eqalign{
 V(\Sigma,H)&=V_1(\Sigma)+V_2(H)+V_3(\Sigma,H)\cr
 &={\rm tr}{2\alpha^4\over g^2}\{\,(\,\Sigma(x)+\Sigma_0\,)^2 +
             M(\,\Sigma(x)+\Sigma_0\,)\,-{1\over 5}{\rm Tr}\Sigma_0^2\,\}^2 \cr
    &+     {2\alpha^4\over g'^2}|\,(\,H(x)+H_0\,)^\dagger(\,H(x)+H_0\,) -
             \mu^2\,|^2, \cr
    &+2\alpha^4({1\over g^2}+{1\over g'^2})|\,(\,\Sigma(x)+\Sigma_0+3M\,)\,
                (\,H(x)+H_0\,)\,|^2. \cr
}\eqno(82)
$$
\par
Eqs.(75) and (82) are essentially same as that of \cham [6].
However, there are  considerably differences how to introduce those equations.
For example, the expression of the auxiliary field $Y_{12}(x)$
in this paper is different from the corresponding auxiliary field $Y_1$
in their paper,
  which difference is owing to the formalism itself. The generation
 mixing matrix $K_{ij}$ in their formalism never appear in ours, which is also
due to the difference of two formulations. The necessity of introducing
the generation matrix is inevitable to the meaningful Higgs potentials
in their formalism. In contrast to this, our formalism is independent of the
 number of generations and meaningful even in one generation.
 In spite of these circumstances, it is interesting
 that \ymh \hskip -1.5mm s in two approaches essentially
 coincide with each others.\par
Eq.(82) contains gauge invariant interaction term (the last term in Eq.(82))
between two Higgs fields
which makes the $SU(3)_c$ color triplet Higgs fields
massive of order of the grand unification scale $M$, whereas it leaves
the usual Higgs doublet massless. Although this is valid only in the tree level
, we should make it in mind to investigate the quantum effects in the
non-commutative geometry on the discrete space. \par
\noindent
\centerline{\bf B.\ fermionic sector}
\par
${\underline 5}$ and ${\underline {10}}$ representations are usually used
in $SU(5)$ Gut to describe the fifteen fermions. We also use
these two representations to incorporate the fermionic sector into our scheme.
Let us begin with writing these two representations. we abbreviate the argument
$x$ in the equations for simplicity hereafter.
${\underline {10}}$ representation is denoted as
$$
    \psi_{10}=\left(\matrix{
              0 & u_3^c &-u_2^c & u_1 & d_1 \cr
             -u_3^c & 0 & u_1^c & u_2 &d_2  \cr
             u_2^c  & -u_1^c  & 0 & u_3 &d_3\cr
             -u_1 &-u_2 &-u_3 & 0 & e^c     \cr
             -d_1 &-d_2 & -d_3 & -e^c  &0     \cr}
             \right)_L            \eqno(83)
$$
which are of course left-handed chiral spinors.
${\underline 5}$ representation is described as
$$
    \psi_{5}=\left(\matrix{
              d_1 \cr
              d_2 \cr
              d_3 \cr
              e^c \cr
              \nu^c\cr}
              \right)_R    \eqno(84)
$$
which are right-handed chiral spinors.
We discuss only one generation for brevity in this section. It is easy
to incorporate three generations.
\par
Two assignments are necessary to provide masses both up quarks and down quarks
as well as in the standard model in sec I$\!$V. By taking account of
the permutation symmetry ${1 \leftrightarrow 2}$ we first make the following
assignment.
$$
\eqalign{
   \psi(x,1)&=\psi(x,2)={a_1\over \sqrt{2}}\psi_{10}, \cr
   \psi(x,3)&=\psi_5,              \cr}   \eqno(85)
$$
where $\dis{a_1\over \sqrt{2}}$ is for the normalization in the final
expression.
We should be more careful to express the differential representations
$A^f(x,n)$ in Eq.(34) corresponding to this assignment.
$\psi(x,1)$ and $\psi(x,3)$ are transformed against the gauge transformation as
$$
            \psi^g(x,1)=g(x)\otimes g(x) \psi(x,1), \hskip 1cm
            \psi^g(x,3)=g(x) \psi(x,3), \eqno(86)
$$
where $g(x)$ is gauge transformation function belonging to $SU(5)$.
With this in mind,
the covariant spinor one form ${\cal D}\psi(x,n)$ in Eq.(36)
is taken to be
$$
\eqalign{
          {\cal D}\psi(x,1)&={\cal D}\psi(x,2)=
          {a_1\over \sqrt{2}}(\, \partial_\mu+(\,A_\mu\otimes 1
          + 1\otimes A_\mu\,)\,)\,\psi_{10}\,dx^\mu\cr
   &\hskip 1cm +{a_1\over \sqrt{2}}(\,\Sigma\otimes 1+1\otimes \Sigma\,)
   \,\psi_{10}\, \chi_2
   + (\,H\otimes 1\,)\,\psi_5\,\chi_3, \cr
     {\cal D}\psi(x,3)&=(\, \partial_\mu+A_\mu\,)\,\psi_5\,dx^\mu
           +{a_1\over \sqrt{2}}(\,1\otimes H^\dagger\,)\, \psi_{10}\,\chi_1
           +{a_1\over \sqrt{2}}(\,1\otimes H^\dagger\,)\, \psi_{10}\,\chi_2,
      }
          \eqno(87)
$$
which mean in components
$$
\eqalign{
          ({\cal D}\psi(x,1)\,)^{ij}&=({\cal D}\psi(x,2)\,)^{ij}=
           {a_1\over \sqrt{2}}(\,\partial_\mu\psi_{10}^{ij}
           +(\,A_\mu\,)^i_k\,\psi^{kj}_{10}
           +  (\,A_\mu\,)^{j}_k\,\psi_{10}^{ik}\,)\,dx^\mu\cr
   &\hskip 1cm +{a_1\over \sqrt{2}}(\,\,\Sigma^i_k\,\psi_{10}^{kj}
   +\,\Sigma^j_k\,\psi^{ik}_{10}\,)\, \chi_2
      + H^i\,\psi_5^j\,\chi_3, \cr
     ({\cal D}\psi(x,3)\,)^i&=( \,\partial_\mu\,\psi_5^i
       +(\,A_\mu\,)^i_k\,\psi^k_5\,)\,dx^\mu
    +{a_1\over \sqrt{2}} H^\dagger_k\, \psi_{10}^{ki}\,\chi_1
        +{a_1\over \sqrt{2}}H^\dagger_k \,\psi_{10}^{ki}\,\chi_2,
      }
          \eqno(88)
$$
It should be noted that $H$ and $\Sigma$ in Eqs.(87), (88) correspond to
$H(x)+H_0$ and $\Sigma(x)+\Sigma_0$ in the bosonic sector, respectively.
\par
The associated spinor one form is written as
$$
\eqalign{
   {\tilde{\cal D}}\psi(x,1)&={\tilde{\cal D}}\psi(x,2)
              ={a_1\over \sqrt{2}}\,(\,\gamma_\mu \,\psi_{10}\,dx^\mu
              -ic_d\,\psi_{10}\,\chi_2-ic_d\,\psi_{10}\,\chi_3\,)\,, \cr
   {\tilde{\cal D}}\psi(x,3)&=\gamma_\mu\,\psi_{5}\,dx^\mu
              -ic_d\,\psi_{5}\,\chi_1-ic_d\,\psi_{5}\,\chi_2\,, \cr}
\eqno(89)
$$
where $c_d$ is related Yukawa coupling constant and so the down quark mass.
\par
With these considerations we can get the Dirac lagrangian for this assignment.
$$
\eqalign{
         {\cal L}_{\mathop{}_{D}}&=
              ia_1^2\,{\rm tr}\,{\bar \psi}_{10}\gamma^\mu(\,\partial_\mu+
               A_\mu\otimes 1+1+\otimes A_\mu\,)\psi_{10}+
         i\,{\bar \psi}_5\gamma^\mu(\,\partial_\mu+A_\mu\,)\,\psi_5 \cr
        &\hskip 1cm  -g_{\mathop{}_{Y}}{\rm tr}\,(\,{\bar \psi}_{10}H\psi_5
         + {\bar \psi}_{5}H^\dagger\psi_{10})\cr},  \eqno(90)
$$
where Yukawa coupling of Higgs to fermions is written in components as
$$
        {\cal L}_{\mathop{}_{\rm Yukawa}}=
        -g_{\mathop{}_{Y}}\sum_{i,j}\{\,({\bar \psi}_{10})_{ji}\,H^i\,\psi_5^j
    + ({\bar \psi}_{5})_j\, H^\dagger_i\,\psi_{10}^{ij}\,\}\,.  \eqno(91)
$$
Eq.(91) provides masses to down-quark and electron.
\par
Eq.(91) leaves up-quarks massless and therefore we need to prepare
the second assignment.
Since we have no room to incorporate the additional
fermions except for $\psi_{10}$ and $\psi_5$, we have to construct such fermion
in terms of $\psi_{10}$ and $\psi_5$. This requires the introduction of
the epsilon tensor $\epsilon^{ijklm}$ of $SU(5)$. We denote such
fermion to be $\omega$ which is transformed as $\underline{10}^\ast$
representation of $SU(5)$. $\omega$ affords  Yukawa coupling to provide masses
to up-quarks as shown later. In components, $\omega$ is expressed as
$$
    \omega^{ijk}={a_2\over \sqrt{24}}\epsilon^{ijklm}(\psi_{10}^c)_{lm}\,,
           \eqno(92)
$$
where $\psi_{10}^c$ is the charge conjugation of $\psi_{10}$
and so $\omega$ is the right handed fermion.
In terms of $\omega$ and $\psi_{10}$, the second assignment is simply given as
$$
     \psi'(x,1)=\psi'(x,2)=\omega, \hskip 1cm \psi'(x,3)=\psi_{10}.
     \eqno(93)
$$
Eqs.(92) and (93) enables us to express the gauge transformation property
of $\psi'(x,n)$.
$$
            \psi'^g(x,1)=g(x)\otimes g(x)\otimes g(x) \psi'(x,1), \hskip 1cm
            \psi'^g(x,3)=g(x)\otimes g(x) \psi'(x,3), \eqno(94)
$$
where $g(x)$ is gauge transformation function belonging to $SU(5)$.
With these considerations, we can write the covariant spinor one form
in Eq.(36) to be
$$
\eqalign{
          {\cal D}\psi'(x,1)&={\cal D}\psi'(x,2)=
          [\, \partial_\mu+(\,A_\mu\otimes 1\otimes 1
  + 1\otimes A_\mu\otimes 1\,+1\otimes 1\otimes A_\mu\,)\,]\,\omega\,dx^\mu\cr
   &+(\,\Sigma\otimes 1\otimes 1\,+1\otimes\Sigma\otimes 1
   +\,1\otimes 1\otimes\Sigma \,)\,\omega\, \chi_2
   + (\,H\otimes 1\otimes 1\,)\,\psi_{10}\,\chi_3, \cr
     {\cal D}\psi'(x,3)&=(\, \partial_\mu+A_\mu\otimes 1
     +1\otimes A_\mu\,)\,\psi_{10}\,dx^\mu \cr
      &\hskip 2cm     +(\,1\otimes 1\otimes H^\dagger\,)\, \omega\,\chi_1
               +(\,1\otimes1\otimes H^\dagger\,)\, \omega\,\chi_2,
      }
          \eqno(95)
$$
which mean in components
$$
\eqalign{
          ({\cal D}\psi'(x,1)\,)^{ijk}&=({\cal D}\psi'(x,2)\,)^{ijk}\cr
          &=           [\,\partial_\mu\omega^{ijk}
           +(\,A_\mu\,)^i_l\,\omega^{ljk}
           +  (\,A_\mu\,)^{j}_l\,\omega^{ilk}\,
           +(\,A_\mu\,)^{k}_l\,\omega^{ijl}\,]\,dx^\mu\cr
   &+(\,\Sigma^i_l\,\omega^{ljk}+\Sigma_l^{j}\,\omega^{ilk}\,
           +\Sigma_l^{k}\,\omega^{ijl}\,)\, \chi_2
      + H^i\,\psi_{10}^{jk}\,\chi_3, \cr
     ({\cal D}\psi'(x,3)\,)^{ij}&=[ \,\partial_\mu\,\psi_{10}^{ij}
       +(\,A_\mu\,)^i_l\,\psi_{10}^{lj}
       +(\,A_\mu\,)^j_l\,\psi_{10}^{il}\,]\,dx^\mu
    + H^\dagger_l\, \omega^{lij}\,\chi_1
        +H^\dagger_l \,\omega^{lij}\,\chi_2,
      }
          \eqno(96)
$$
The associated spinor one form is written as
$$
\eqalign{
   {\tilde{\cal D}}\psi'(x,1)&={\tilde{\cal D}}\psi'(x,2)
              =(\,\gamma_\mu \,\omega\,dx^\mu
              -ic_u\,\omega\,\chi_2-ic_u\,\omega\,\chi_3\,)\,, \cr
   {\tilde{\cal D}}\psi'(x,3)&=\gamma_\mu\,\psi_{10}\,dx^\mu
              -ic_u\,\psi_{10}\,\chi_1-ic_u\,\psi_{10}\,\chi_2\,, \cr}
\eqno(97)
$$
where $c_u$ is related Yukawa coupling constant and so the up quark mass.
\par
With these considerations we can get the Dirac lagrangian for this assignment.
$$
\eqalign{
    {\cal L}'_{\mathop{}_{D}}&=i(\,a_2^2+a_3^2\,)\,{\rm tr}\,
         {\bar \psi}_{10}\,\gamma^\mu\,
         (\,\partial_\mu+
               A_\mu\otimes 1+1+\otimes A_\mu\,)\,\psi_{10}\cr
        &\hskip 1cm   -g'_{\mathop{}_{Y}}\,
        {\rm tr}\,(\,{\bar \omega}\,(\,H\,\otimes\psi_{10})
         + ({\bar \psi}_{10}\,\otimes H^\dagger\,)\,\omega\,),\cr}  \eqno(98)
$$
where Yukawa coupling of Higgs to fermions is written in components as
$$
        {\cal L}'_{\mathop{}_{\rm Yukawa}}=
        -g'_{\mathop{}_{Y}}\sum_{i,j,k,l,m}\,
        [\,\epsilon_{ijklm}\,({\bar \psi^c}_{10})^{ij}\,H^k\,\psi_{10}^{lm}
    + \epsilon^{ijklm}\,(\,{\bar \psi}_{10}\,)_{ij}\, H^\dagger_k\,
    (\,\psi_{10}^c\,)_{lm}\,]\,,
      \eqno(99)
$$
with $g'_{\mathop{}_{Y}}$  appropriately chosen coupling constant.\hskip -0.1mm
{}From Eq.(98) together with Eq.(90) the kinetic term of $\psi_{10}$ is
 normalized if $a_1^2+a_2^2+a_3^2=1$ is satisfied. Up-quarks become massive
 owing to the Yukawa coupling in Eq.(99).
\par

\section{ \hskip 4.1cm {\bf V\hskip -0.4mm I. Conclusions and discussions}}
We have proposed the general framework to construct the spontaneously
broken gauge theory not only in the bosonic sector but also
in the fermionic sector. The standard model and $SU(5)$ Gut are
reconstructed according to this framework.
Our formalism is based on the generalized differential
calculus which is more familiar than Connes's original approach.
It should be stressed that the method to introduce the Dirac lagrangian for
fermions is original, not in any text book of differential geometry.
\par
Although \cham make Connes's approach more understandable
and also applicable to Gut,
its mathematical settings are still difficult to prevent us from
true understanding. Our differential calculus characterized by Eqs.(2) and (8)
together with the sifting rule is an extention of the ordinary differential
calculus, so that the mathematical settings become rather easy.
The largest difference between ours and theirs in the resulted equations
is the existence of the generation matrix $K$. Our formalism is independent
of this factor, and so meaningful even in one generation.
Although the non-commutative approach enables us to predict Higgs mass
in tree level, the existence of $K$ makes this predictive power invalid.
However, the unified picture of the gauge field and Higgs field
as the generalized  connection in non-commutative geometry is excellent,
which point of view is due to Connes and \cham.
\par
Our starting point in this paper is Eq.(1), where the unidentified factor
$a_i(x,n)$ appears. The matrix $M_{nm}$ appears through the generalized
differential derivative. The gauge field and Higgs field are written
in terms of $a_i(x,n)$ and $M_{nm}$ in Eq.(13). We again quote it.
$$
  \eqalign{
    A_\mu(x,n) &\= \sum_{i}a_{i}^\dagger(x,n)\partial_\mu a_{i}(x,n), \cr
\Phi_{nm}(x) &\= \sum_{i}a_{i}^\dagger(x,n)\,(-a_i(x,n)M_{nm}
            + M_{nm}a_i(x,m)),\cr
}  \eqno(100)
$$
which makes us imagine that $a_i(x,n)$ would be the more fundamental object
and the gauge and Higgs bosons were composed of them. In addition to this,
$a_i(x,n)$ is restricted to be $\sum_i a_i^\dagger(x,n)a_i(x,n)=1$ which
seems to be the normalization condition for the field.
It should be remarked whether $a_i(x,n)$ is the physical object
experimentally observed in future or not. \par
We have introduced the parameter $\delta$ in Eqs.(54)$\sim$(56) which is the
ratio of the coupling constants $g_2$ and $g_3$. $\delta$ is usually taken
to be 1 in non-commutative geometry, whereas we let it free parameter.
With respect to $\delta$,
we take the strategy [9] that $g_2$ and $g_3$ coincide
with each other at the energy $E_{\mathop{}_{WS}}$,
at which the electromagnetic and weak interactions are unified.
{}From Eq.(56), $\sin \theta_{\mathop{}_{W}}=\dis{2(1-r)^2\over4-4r+3r^2}$
at $E_{\mathop{}_{WS}}$. According to the renormalization equation,
the running coupling constants depend on the energy $E$ at which they are
measured and $\delta$ shifts from 1.  Thus, we take account of
a part of the quantum effects in this context. However, it is questionable
whether Eqs.(57) and (58) are stable or not against the quantum effects as
pointed out by ${\grave {\rm A}}$lvarez et al. [11].
This problem still remains unsolved in non-commutative geometry approach.
\par
$SU(5)$ Gut has been ruled out as a true theory in particle physics due to the
accurate experiments by LEP and no-detection of proton decay.
The modified $SO(10)$ unified model and MSSM are more promising theories
nowadays. $SO(10)$ model is also in the scope of our formalism.

\vskip 0.0cm
\par
\centerline{\bf Acknowledgements}
\par
The author would like to
express his sincere thanks to
Professor M.~Morita and J.~Iizuka
for useful suggestion on the noncommutative geometry
and invaluable discussions and
Professors H.~Kase
and
M.~Tanaka for informing some references and
useful discussions.
\vskip 2cm

\centerline{{\bf References}}

\def\pl{Phys. Lett. }
\def\np{Nucl. Phys.}

\def\prl{Phys. Rev. Lett.}
\def\pr{Phys. Rev. D}
\def\mp{Int. Journ. Mod. Phys. }

\item{[1]} S.~Weinberg, {\prl} {\bf 19} 1264(1967). \hfill\break
A.~Salam, {\it Weak and Electromagnetic Interactions, Elementary
Particle Theory},
ed. N.~Svartholm (John Wiley and Sons, Inc., New York, London,
Sydney, 1968), p.367.
\item{[2]}
H.~Georgi and S.L.~Glashow, \prl {\bf 32} (1974) 438, \hfill\break
C.~Pati and A.~ Salam, \pr {\bf 8} (1973) 1240, \hfill\break
for a review of Gut, see G.Ross, Grand unified theories, (Frontiers in Physics
Series, Vol.{\bf 60} (Benjamin, New York) and references therein.
\item{[3]}
K.~Yamawaki, Proceeding of 1989 Workshop on Dynamical Symmetry
Breaking at Nagoya, p.11.

\item{[4]}A.~Connes,
p.9 in {\it The Interface of Mathematics and Particle
Physics}, ed. D.~G.~Quillen, G.~B.~Segal, and Tsou.~S.~T.,
Clarendon Press, Oxford, 1990. See also,
Alain Connes and J. Lott,
Nucl. Phys. {\bf B}(Proc. Suppl.) {\bf 18B} (1990) 57.
\item{[5]} D.~Kastler,$\;$
Introduction to Alain Connes' Non-Commutative
Differential Geometry, CPT-86/PE.1929. \hfill\break
See also,
D.~Kastler,$\;$
Introduction to non-commutative geometry and Yang-Mills
model-building,
XIXth. International Conference on Differential Geometric Methods
in Theoretical Physics. Rapallo (Italy) June 1990. Springer Lect. Notes
in Phys.$\;$ {\bf 375}(1990) 25-44 \hfill\break
R. Coquereaux, G. Esposito-Farese and G. Vaillant,
Nucl. Phys. {\bf B353} (1991) 689;\hfill\break
R.~Coquereaux, R.~H{\"a}{\ss}ling, N.~A.~Papadopoulos,
and F.~Scheck,
{\mp} {\bf A7} (1992) 2809.
\item{[6]} A.~H.~Chamseddine, G.~Felder and J.~Fr\"olich,
\pl$\;$ {\bf B296} (1992) 109; \ \np$\;$ B395 (1993) 672.\hfill\break
See also, A.~H.~Chamseddine and J.~Fr\"olich,
^^ ^^  SO(10) Unification in Non-Commutative Geometry ",
ZU-TH-10/1993, ETH/ \ TH/93-12, 12 March 1992.

\item{[7]} A.~Sitarz, \pl,\ {\bf B308}(1993) 311.                 \hfill\break
See also, Jagiellonian Univ. preprint,TPJU$-$7/92
^^ ^^  Non-commutative Geometry and Gauge Theory on Discrete Groups ".
\hfill\break
H-G.~Ding, H-Y.~Gou, J-M.~Li and K.~Wu,
preprint, ASITP-93-23, CCAST-93-5,
^^ ^^  Higgs as Gauge Fields on Discrete Groups and Standard Models
for Electroweak and Electroweak-Strong Interactions ".
\hfill\break
 This Beijing group paper
 also reconstructed \ws by use of Sitarz' formalism
 developed in the above preprint.
 The authors are thankful to Professor T.~Saito for informing
 this reference to us.
\item{[8]} K.~Morita and Y.~Okumura, Nagoya Univ. preprint, DPNU-93-25,
\hfill\break
^^ ^^  Weinberg-Salam Theory in Non-Commutative Geometry ",
\item{[9]} K.~Morita and Y.~Okumura, Nagoya Univ. preprint, DPNU-93-38,
\hfill\break
^^ ^^   Reconstruction of \ws in non-commutative geometry on $M_4\times Z_2$ ",
\item{[10]} K.~Morita and Y.~Okumura, Nogoya Univ. preprint, DPNU-93-39
\hfill\break
$B@(J^^ ^^  Gauge Theory on Discrete Space $M_4\times Z_{\mathop{}_{N}}$ and
SU(5) Gut
in non-commutative Geometry ".
\item{[11]} E.~\'Alvarez, J.~M.~Gracia-Bond\'ia and
C.~P.~Martin, \pl$\;$ {\bf B306}(1993) 55.
\bye